
\catcode`\@=11


\message{Loading jyTeX fonts...}



\font\vptrm=cmr5 \font\vptmit=cmmi5 \font\vptsy=cmsy5 \font\vptbf=cmbx5

\skewchar\vptmit='177 \skewchar\vptsy='60 \fontdimen16
\vptsy=\the\fontdimen17 \vptsy

\def\vpt{\ifmmode\err@badsizechange\else
     \@mathfontinit
     \textfont0=\vptrm  \scriptfont0=\vptrm  \scriptscriptfont0=\vptrm
     \textfont1=\vptmit \scriptfont1=\vptmit \scriptscriptfont1=\vptmit
     \textfont2=\vptsy  \scriptfont2=\vptsy  \scriptscriptfont2=\vptsy
     \textfont3=\xptex  \scriptfont3=\xptex  \scriptscriptfont3=\xptex
     \textfont\bffam=\vptbf
     \scriptfont\bffam=\vptbf
     \scriptscriptfont\bffam=\vptbf
     \@fontstyleinit
     \def\rm{\vptrm\fam=\z@}%
     \def\bf{\vptbf\fam=\bffam}%
     \def\oldstyle{\vptmit\fam=\@ne}%
     \rm\fi}


\font\viptrm=cmr6 \font\viptmit=cmmi6 \font\viptsy=cmsy6
\font\viptbf=cmbx6

\skewchar\viptmit='177 \skewchar\viptsy='60 \fontdimen16
\viptsy=\the\fontdimen17 \viptsy

\def\vipt{\ifmmode\err@badsizechange\else
     \@mathfontinit
     \textfont0=\viptrm  \scriptfont0=\vptrm  \scriptscriptfont0=\vptrm
     \textfont1=\viptmit \scriptfont1=\vptmit \scriptscriptfont1=\vptmit
     \textfont2=\viptsy  \scriptfont2=\vptsy  \scriptscriptfont2=\vptsy
     \textfont3=\xptex   \scriptfont3=\xptex  \scriptscriptfont3=\xptex
     \textfont\bffam=\viptbf
     \scriptfont\bffam=\vptbf
     \scriptscriptfont\bffam=\vptbf
     \@fontstyleinit
     \def\rm{\viptrm\fam=\z@}%
     \def\bf{\viptbf\fam=\bffam}%
     \def\oldstyle{\viptmit\fam=\@ne}%
     \rm\fi}

\font\viiptrm=cmr7 \font\viiptmit=cmmi7 \font\viiptsy=cmsy7
\font\viiptit=cmti7 \font\viiptbf=cmbx7

\skewchar\viiptmit='177 \skewchar\viiptsy='60 \fontdimen16
\viiptsy=\the\fontdimen17 \viiptsy

\def\viipt{\ifmmode\err@badsizechange\else
     \@mathfontinit
     \textfont0=\viiptrm  \scriptfont0=\vptrm  \scriptscriptfont0=\vptrm
     \textfont1=\viiptmit \scriptfont1=\vptmit \scriptscriptfont1=\vptmit
     \textfont2=\viiptsy  \scriptfont2=\vptsy  \scriptscriptfont2=\vptsy
     \textfont3=\xptex    \scriptfont3=\xptex  \scriptscriptfont3=\xptex
     \textfont\itfam=\viiptit
     \scriptfont\itfam=\viiptit
     \scriptscriptfont\itfam=\viiptit
     \textfont\bffam=\viiptbf
     \scriptfont\bffam=\vptbf
     \scriptscriptfont\bffam=\vptbf
     \@fontstyleinit
     \def\rm{\viiptrm\fam=\z@}%
     \def\it{\viiptit\fam=\itfam}%
     \def\bf{\viiptbf\fam=\bffam}%
     \def\oldstyle{\viiptmit\fam=\@ne}%
     \rm\fi}


\font\viiiptrm=cmr8 \font\viiiptmit=cmmi8 \font\viiiptsy=cmsy8
\font\viiiptit=cmti8
\font\viiiptbf=cmbx8

\skewchar\viiiptmit='177 \skewchar\viiiptsy='60 \fontdimen16
\viiiptsy=\the\fontdimen17 \viiiptsy

\def\viiipt{\ifmmode\err@badsizechange\else
     \@mathfontinit
     \textfont0=\viiiptrm  \scriptfont0=\viptrm  \scriptscriptfont0=\vptrm
     \textfont1=\viiiptmit \scriptfont1=\viptmit \scriptscriptfont1=\vptmit
     \textfont2=\viiiptsy  \scriptfont2=\viptsy  \scriptscriptfont2=\vptsy
     \textfont3=\xptex     \scriptfont3=\xptex   \scriptscriptfont3=\xptex
     \textfont\itfam=\viiiptit
     \scriptfont\itfam=\viiptit
     \scriptscriptfont\itfam=\viiptit
     \textfont\bffam=\viiiptbf
     \scriptfont\bffam=\viptbf
     \scriptscriptfont\bffam=\vptbf
     \@fontstyleinit
     \def\rm{\viiiptrm\fam=\z@}%
     \def\it{\viiiptit\fam=\itfam}%
     \def\bf{\viiiptbf\fam=\bffam}%
     \def\oldstyle{\viiiptmit\fam=\@ne}%
     \rm\fi}


\def\getixpt{%
     \font\ixptrm=cmr9
     \font\ixptmit=cmmi9
     \font\ixptsy=cmsy9
     \font\ixptit=cmti9
     \font\ixptbf=cmbx9
     \skewchar\ixptmit='177 \skewchar\ixptsy='60
     \fontdimen16 \ixptsy=\the\fontdimen17 \ixptsy}

\def\ixpt{\ifmmode\err@badsizechange\else
     \@mathfontinit
     \textfont0=\ixptrm  \scriptfont0=\viiptrm  \scriptscriptfont0=\vptrm
     \textfont1=\ixptmit \scriptfont1=\viiptmit \scriptscriptfont1=\vptmit
     \textfont2=\ixptsy  \scriptfont2=\viiptsy  \scriptscriptfont2=\vptsy
     \textfont3=\xptex   \scriptfont3=\xptex    \scriptscriptfont3=\xptex
     \textfont\itfam=\ixptit
     \scriptfont\itfam=\viiptit
     \scriptscriptfont\itfam=\viiptit
     \textfont\bffam=\ixptbf
     \scriptfont\bffam=\viiptbf
     \scriptscriptfont\bffam=\vptbf
     \@fontstyleinit
     \def\rm{\ixptrm\fam=\z@}%
     \def\it{\ixptit\fam=\itfam}%
     \def\bf{\ixptbf\fam=\bffam}%
     \def\oldstyle{\ixptmit\fam=\@ne}%
     \rm\fi}


\font\xptrm=cmr10 \font\xptmit=cmmi10 \font\xptsy=cmsy10
\font\xptex=cmex10 \font\xptit=cmti10 \font\xptsl=cmsl10
\font\xptbf=cmbx10 \font\xpttt=cmtt10 \font\xptss=cmss10
\font\xptsc=cmcsc10 \font\xptbfs=cmb10 \font\xptbmit=cmmib10

\skewchar\xptmit='177 \skewchar\xptbmit='177 \skewchar\xptsy='60
\fontdimen16 \xptsy=\the\fontdimen17 \xptsy

\def\xpt{\ifmmode\err@badsizechange\else
     \@mathfontinit
     \textfont0=\xptrm  \scriptfont0=\viiptrm  \scriptscriptfont0=\vptrm
     \textfont1=\xptmit \scriptfont1=\viiptmit \scriptscriptfont1=\vptmit
     \textfont2=\xptsy  \scriptfont2=\viiptsy  \scriptscriptfont2=\vptsy
     \textfont3=\xptex  \scriptfont3=\xptex    \scriptscriptfont3=\xptex
     \textfont\itfam=\xptit
     \scriptfont\itfam=\viiptit
     \scriptscriptfont\itfam=\viiptit
     \textfont\bffam=\xptbf
     \scriptfont\bffam=\viiptbf
     \scriptscriptfont\bffam=\vptbf
     \textfont\bfsfam=\xptbfs
     \scriptfont\bfsfam=\viiptbf
     \scriptscriptfont\bfsfam=\vptbf
     \textfont\bmitfam=\xptbmit
     \scriptfont\bmitfam=\viiptmit
     \scriptscriptfont\bmitfam=\vptmit
     \@fontstyleinit
     \def\rm{\xptrm\fam=\z@}%
     \def\it{\xptit\fam=\itfam}%
     \def\sl{\xptsl}%
     \def\bf{\xptbf\fam=\bffam}%
     \def\tt{\xpttt}%
     \def\ss{\xptss}%
     \def\sc{\xptsc}%
     \def\bfs{\xptbfs\fam=\bfsfam}%
     \def\bmit{\fam=\bmitfam}%
     \def\oldstyle{\xptmit\fam=\@ne}%
     \rm\fi}


\def\getxipt{%
     \font\xiptrm=cmr10  scaled\magstephalf
     \font\xiptmit=cmmi10 scaled\magstephalf
     \font\xiptsy=cmsy10 scaled\magstephalf
     \font\xiptex=cmex10 scaled\magstephalf
     \font\xiptit=cmti10 scaled\magstephalf
     \font\xiptsl=cmsl10 scaled\magstephalf
     \font\xiptbf=cmbx10 scaled\magstephalf
     \font\xipttt=cmtt10 scaled\magstephalf
     \font\xiptss=cmss10 scaled\magstephalf
     \skewchar\xiptmit='177 \skewchar\xiptsy='60
     \fontdimen16 \xiptsy=\the\fontdimen17 \xiptsy}

\def\xipt{\ifmmode\err@badsizechange\else
     \@mathfontinit
     \textfont0=\xiptrm  \scriptfont0=\viiiptrm  \scriptscriptfont0=\viptrm
     \textfont1=\xiptmit \scriptfont1=\viiiptmit \scriptscriptfont1=\viptmit
     \textfont2=\xiptsy  \scriptfont2=\viiiptsy  \scriptscriptfont2=\viptsy
     \textfont3=\xiptex  \scriptfont3=\xptex     \scriptscriptfont3=\xptex
     \textfont\itfam=\xiptit
     \scriptfont\itfam=\viiiptit
     \scriptscriptfont\itfam=\viiptit
     \textfont\bffam=\xiptbf
     \scriptfont\bffam=\viiiptbf
     \scriptscriptfont\bffam=\viptbf
     \@fontstyleinit
     \def\rm{\xiptrm\fam=\z@}%
     \def\it{\xiptit\fam=\itfam}%
     \def\sl{\xiptsl}%
     \def\bf{\xiptbf\fam=\bffam}%
     \def\tt{\xipttt}%
     \def\ss{\xiptss}%
     \def\oldstyle{\xiptmit\fam=\@ne}%
     \rm\fi}


\font\xiiptrm=cmr12 \font\xiiptmit=cmmi12 \font\xiiptsy=cmsy10
scaled\magstep1 \font\xiiptex=cmex10  scaled\magstep1
\font\xiiptit=cmti12 \font\xiiptsl=cmsl12 \font\xiiptbf=cmbx12
\font\xiiptss=cmss12 \font\xiiptsc=cmcsc10 scaled\magstep1
\font\xiiptbfs=cmb10  scaled\magstep1 \font\xiiptbmit=cmmib10
scaled\magstep1

\skewchar\xiiptmit='177 \skewchar\xiiptbmit='177 \skewchar\xiiptsy='60
\fontdimen16 \xiiptsy=\the\fontdimen17 \xiiptsy

\def\xiipt{\ifmmode\err@badsizechange\else
     \@mathfontinit
     \textfont0=\xiiptrm  \scriptfont0=\viiiptrm  \scriptscriptfont0=\viptrm
     \textfont1=\xiiptmit \scriptfont1=\viiiptmit \scriptscriptfont1=\viptmit
     \textfont2=\xiiptsy  \scriptfont2=\viiiptsy  \scriptscriptfont2=\viptsy
     \textfont3=\xiiptex  \scriptfont3=\xptex     \scriptscriptfont3=\xptex
     \textfont\itfam=\xiiptit
     \scriptfont\itfam=\viiiptit
     \scriptscriptfont\itfam=\viiptit
     \textfont\bffam=\xiiptbf
     \scriptfont\bffam=\viiiptbf
     \scriptscriptfont\bffam=\viptbf
     \textfont\bfsfam=\xiiptbfs
     \scriptfont\bfsfam=\viiiptbf
     \scriptscriptfont\bfsfam=\viptbf
     \textfont\bmitfam=\xiiptbmit
     \scriptfont\bmitfam=\viiiptmit
     \scriptscriptfont\bmitfam=\viptmit
     \@fontstyleinit
     \def\rm{\xiiptrm\fam=\z@}%
     \def\it{\xiiptit\fam=\itfam}%
     \def\sl{\xiiptsl}%
     \def\bf{\xiiptbf\fam=\bffam}%
     \def\tt{\xiipttt}%
     \def\ss{\xiiptss}%
     \def\sc{\xiiptsc}%
     \def\bfs{\xiiptbfs\fam=\bfsfam}%
     \def\bmit{\fam=\bmitfam}%
     \def\oldstyle{\xiiptmit\fam=\@ne}%
     \rm\fi}


\def\getxiiipt{%
     \font\xiiiptrm=cmr12  scaled\magstephalf
     \font\xiiiptmit=cmmi12 scaled\magstephalf
     \font\xiiiptsy=cmsy9  scaled\magstep2
     \font\xiiiptit=cmti12 scaled\magstephalf
     \font\xiiiptsl=cmsl12 scaled\magstephalf
     \font\xiiiptbf=cmbx12 scaled\magstephalf
     \font\xiiipttt=cmtt12 scaled\magstephalf
     \font\xiiiptss=cmss12 scaled\magstephalf
     \skewchar\xiiiptmit='177 \skewchar\xiiiptsy='60
     \fontdimen16 \xiiiptsy=\the\fontdimen17 \xiiiptsy}

\def\xiiipt{\ifmmode\err@badsizechange\else
     \@mathfontinit
     \textfont0=\xiiiptrm  \scriptfont0=\xptrm  \scriptscriptfont0=\viiptrm
     \textfont1=\xiiiptmit \scriptfont1=\xptmit \scriptscriptfont1=\viiptmit
     \textfont2=\xiiiptsy  \scriptfont2=\xptsy  \scriptscriptfont2=\viiptsy
     \textfont3=\xivptex   \scriptfont3=\xptex  \scriptscriptfont3=\xptex
     \textfont\itfam=\xiiiptit
     \scriptfont\itfam=\xptit
     \scriptscriptfont\itfam=\viiptit
     \textfont\bffam=\xiiiptbf
     \scriptfont\bffam=\xptbf
     \scriptscriptfont\bffam=\viiptbf
     \@fontstyleinit
     \def\rm{\xiiiptrm\fam=\z@}%
     \def\it{\xiiiptit\fam=\itfam}%
     \def\sl{\xiiiptsl}%
     \def\bf{\xiiiptbf\fam=\bffam}%
     \def\tt{\xiiipttt}%
     \def\ss{\xiiiptss}%
     \def\oldstyle{\xiiiptmit\fam=\@ne}%
     \rm\fi}


\font\xivptrm=cmr12   scaled\magstep1 \font\xivptmit=cmmi12
scaled\magstep1 \font\xivptsy=cmsy10  scaled\magstep2
\font\xivptex=cmex10  scaled\magstep2 \font\xivptit=cmti12
scaled\magstep1 \font\xivptsl=cmsl12  scaled\magstep1
\font\xivptbf=cmbx12  scaled\magstep1
\font\xivptss=cmss12  scaled\magstep1 \font\xivptsc=cmcsc10
scaled\magstep2 \font\xivptbfs=cmb10  scaled\magstep2
\font\xivptbmit=cmmib10 scaled\magstep2

\skewchar\xivptmit='177 \skewchar\xivptbmit='177 \skewchar\xivptsy='60
\fontdimen16 \xivptsy=\the\fontdimen17 \xivptsy

\def\xivpt{\ifmmode\err@badsizechange\else
     \@mathfontinit
     \textfont0=\xivptrm  \scriptfont0=\xptrm  \scriptscriptfont0=\viiptrm
     \textfont1=\xivptmit \scriptfont1=\xptmit \scriptscriptfont1=\viiptmit
     \textfont2=\xivptsy  \scriptfont2=\xptsy  \scriptscriptfont2=\viiptsy
     \textfont3=\xivptex  \scriptfont3=\xptex  \scriptscriptfont3=\xptex
     \textfont\itfam=\xivptit
     \scriptfont\itfam=\xptit
     \scriptscriptfont\itfam=\viiptit
     \textfont\bffam=\xivptbf
     \scriptfont\bffam=\xptbf
     \scriptscriptfont\bffam=\viiptbf
     \textfont\bfsfam=\xivptbfs
     \scriptfont\bfsfam=\xptbfs
     \scriptscriptfont\bfsfam=\viiptbf
     \textfont\bmitfam=\xivptbmit
     \scriptfont\bmitfam=\xptbmit
     \scriptscriptfont\bmitfam=\viiptmit
     \@fontstyleinit
     \def\rm{\xivptrm\fam=\z@}%
     \def\it{\xivptit\fam=\itfam}%
     \def\sl{\xivptsl}%
     \def\bf{\xivptbf\fam=\bffam}%
     \def\tt{\xivpttt}%
     \def\ss{\xivptss}%
     \def\sc{\xivptsc}%
     \def\bfs{\xivptbfs\fam=\bfsfam}%
     \def\bmit{\fam=\bmitfam}%
     \def\oldstyle{\xivptmit\fam=\@ne}%
     \rm\fi}


\font\xviiptrm=cmr17 \font\xviiptmit=cmmi12 scaled\magstep2
\font\xviiptsy=cmsy10 scaled\magstep3 \font\xviiptex=cmex10
scaled\magstep3 \font\xviiptit=cmti12 scaled\magstep2
\font\xviiptbf=cmbx12 scaled\magstep2 \font\xviiptbfs=cmb10
scaled\magstep3

\skewchar\xviiptmit='177 \skewchar\xviiptsy='60 \fontdimen16
\xviiptsy=\the\fontdimen17 \xviiptsy

\def\xviipt{\ifmmode\err@badsizechange\else
     \@mathfontinit
     \textfont0=\xviiptrm  \scriptfont0=\xiiptrm  \scriptscriptfont0=\viiiptrm
     \textfont1=\xviiptmit \scriptfont1=\xiiptmit \scriptscriptfont1=\viiiptmit
     \textfont2=\xviiptsy  \scriptfont2=\xiiptsy  \scriptscriptfont2=\viiiptsy
     \textfont3=\xviiptex  \scriptfont3=\xiiptex  \scriptscriptfont3=\xptex
     \textfont\itfam=\xviiptit
     \scriptfont\itfam=\xiiptit
     \scriptscriptfont\itfam=\viiiptit
     \textfont\bffam=\xviiptbf
     \scriptfont\bffam=\xiiptbf
     \scriptscriptfont\bffam=\viiiptbf
     \textfont\bfsfam=\xviiptbfs
     \scriptfont\bfsfam=\xiiptbfs
     \scriptscriptfont\bfsfam=\viiiptbf
     \@fontstyleinit
     \def\rm{\xviiptrm\fam=\z@}%
     \def\it{\xviiptit\fam=\itfam}%
     \def\bf{\xviiptbf\fam=\bffam}%
     \def\bfs{\xviiptbfs\fam=\bfsfam}%
     \def\oldstyle{\xviiptmit\fam=\@ne}%
     \rm\fi}


\font\xxiptrm=cmr17  scaled\magstep1


\def\xxipt{\ifmmode\err@badsizechange\else
     \@mathfontinit
     \@fontstyleinit
     \def\rm{\xxiptrm\fam=\z@}%
     \rm\fi}


\font\xxvptrm=cmr17  scaled\magstep2


\def\xxvpt{\ifmmode\err@badsizechange\else
     \@mathfontinit
     \@fontstyleinit
     \def\rm{\xxvptrm\fam=\z@}%
     \rm\fi}




\message{Loading jyTeX macros...}

\message{modifications to plain.tex,}


\def\newcount{\alloc@0\count\countdef\insc@unt}
\def\newdimen{\alloc@1\dimen\dimendef\insc@unt}
\def\newskip{\alloc@2\skip\skipdef\insc@unt}
\def\newmuskip{\alloc@3\muskip\muskipdef\@cclvi}
\def\newbox{\alloc@4\box\chardef\insc@unt}
\def\newtoks{\alloc@5\toks\toksdef\@cclvi}
\def\newhelp#1#2{\newtoks#1\global#1\expandafter{\csname#2\endcsname}}
\def\newread{\alloc@6\read\chardef\sixt@@n}
\def\newwrite{\alloc@7\write\chardef\sixt@@n}
\def\newfam{\alloc@8\fam\chardef\sixt@@n}
\def\newinsert#1{\global\advance\insc@unt by\m@ne
     \ch@ck0\insc@unt\count
     \ch@ck1\insc@unt\dimen
     \ch@ck2\insc@unt\skip
     \ch@ck4\insc@unt\box
     \allocationnumber=\insc@unt
     \global\chardef#1=\allocationnumber
     \wlog{\string#1=\string\insert\the\allocationnumber}}
\def\newif#1{\count@\escapechar \escapechar\m@ne
     \expandafter\expandafter\expandafter
          \xdef\@if#1{true}{\let\noexpand#1=\noexpand\iftrue}%
     \expandafter\expandafter\expandafter
          \xdef\@if#1{false}{\let\noexpand#1=\noexpand\iffalse}%
     \global\@if#1{false}\escapechar=\count@}


\newlinechar=`\^^J
\overfullrule=0pt




\let\itfam=\undefined

\let\bffam=\undefined

\count18=3


\chardef\sharps="19


\mathchardef\alpha="710B \mathchardef\beta="710C \mathchardef\gamma="710D
\mathchardef\delta="710E \mathchardef\epsilon="710F
\mathchardef\zeta="7110 \mathchardef\eta="7111 \mathchardef\theta="7112
\mathchardef\iota="7113 \mathchardef\kappa="7114
\mathchardef\lambda="7115 \mathchardef\mu="7116 \mathchardef\nu="7117
\mathchardef\xi="7118 \mathchardef\pi="7119 \mathchardef\rho="711A
\mathchardef\sigma="711B \mathchardef\tau="711C
\mathchardef\upsilon="711D \mathchardef\phi="711E \mathchardef\chi="711F
\mathchardef\psi="7120 \mathchardef\omega="7121
\mathchardef\varepsilon="7122 \mathchardef\vartheta="7123
\mathchardef\varpi="7124 \mathchardef\varrho="7125
\mathchardef\varsigma="7126 \mathchardef\varphi="7127
\mathchardef\imath="717B \mathchardef\jmath="717C \mathchardef\ell="7160
\mathchardef\wp="717D \mathchardef\partial="7140 \mathchardef\flat="715B
\mathchardef\natural="715C \mathchardef\sharp="715D



\def\angle{{\vbox{\ialign{$\m@th\scriptstyle##$\crcr
     \not\mathrel{\mkern14mu}\crcr
     \noalign{\nointerlineskip}
     \mkern2.5mu\leaders\hrule height.34\rp@\hfill\mkern2.5mu\crcr}}}}
\def\vdots{\vbox{\baselineskip4\rp@ \lineskiplimit\z@
     \kern6\rp@\hbox{.}\hbox{.}\hbox{.}}}
\def\ddots{\mathinner{\mkern1mu\raise7\rp@\vbox{\kern7\rp@\hbox{.}}\mkern2mu
     \raise4\rp@\hbox{.}\mkern2mu\raise\rp@\hbox{.}\mkern1mu}}
\def\overrightarrow#1{\vbox{\ialign{##\crcr
     \rightarrowfill\crcr
     \noalign{\kern-\rp@\nointerlineskip}
     $\hfil\displaystyle{#1}\hfil$\crcr}}}
\def\overleftarrow#1{\vbox{\ialign{##\crcr
     \leftarrowfill\crcr
     \noalign{\kern-\rp@\nointerlineskip}
     $\hfil\displaystyle{#1}\hfil$\crcr}}}
\def\overbrace#1{\mathop{\vbox{\ialign{##\crcr
     \noalign{\kern3\rp@}
     \downbracefill\crcr
     \noalign{\kern3\rp@\nointerlineskip}
     $\hfil\displaystyle{#1}\hfil$\crcr}}}\limits}
\def\underbrace#1{\mathop{\vtop{\ialign{##\crcr
     $\hfil\displaystyle{#1}\hfil$\crcr
     \noalign{\kern3\rp@\nointerlineskip}
     \upbracefill\crcr
     \noalign{\kern3\rp@}}}}\limits}
\def\big#1{{\hbox{$\left#1\vbox to8.5\rp@ {}\right.\n@space$}}}
\def\Big#1{{\hbox{$\left#1\vbox to11.5\rp@ {}\right.\n@space$}}}
\def\bigg#1{{\hbox{$\left#1\vbox to14.5\rp@ {}\right.\n@space$}}}
\def\Bigg#1{{\hbox{$\left#1\vbox to17.5\rp@ {}\right.\n@space$}}}
\def\@vereq#1#2{\lower.5\rp@\vbox{\baselineskip\z@skip\lineskip-.5\rp@
     \ialign{$\m@th#1\hfil##\hfil$\crcr#2\crcr=\crcr}}}
\def\rlh@#1{\vcenter{\hbox{\ooalign{\raise2\rp@
     \hbox{$#1\rightharpoonup$}\crcr
     $#1\leftharpoondown$}}}}
\def\bordermatrix#1{\begingroup\m@th
     \setbox\z@\vbox{%
          \def\cr{\crcr\noalign{\kern2\rp@\global\let\cr\endline}}%
          \ialign{$##$\hfil\kern2\rp@\kern\p@renwd
               &\thinspace\hfil$##$\hfil&&\quad\hfil$##$\hfil\crcr
               \omit\strut\hfil\crcr
               \noalign{\kern-\baselineskip}%
               #1\crcr\omit\strut\cr}}%
     \setbox\tw@\vbox{\unvcopy\z@\global\setbox\@ne\lastbox}%
     \setbox\tw@\hbox{\unhbox\@ne\unskip\global\setbox\@ne\lastbox}%
     \setbox\tw@\hbox{$\kern\wd\@ne\kern-\p@renwd\left(\kern-\wd\@ne
          \global\setbox\@ne\vbox{\box\@ne\kern2\rp@}%
          \vcenter{\kern-\ht\@ne\unvbox\z@\kern-\baselineskip}%
          \,\right)$}%
     \null\;\vbox{\kern\ht\@ne\box\tw@}\endgroup}
\def\endinsert{\egroup
     \if@mid\dimen@\ht\z@
          \advance\dimen@\dp\z@
          \advance\dimen@12\rp@
          \advance\dimen@\pagetotal
          \ifdim\dimen@>\pagegoal\@midfalse\p@gefalse\fi
     \fi
     \if@mid\bigskip\box\z@
          \bigbreak
     \else\insert\topins{\penalty100 \splittopskip\z@skip
               \splitmaxdepth\maxdimen\floatingpenalty\z@
               \ifp@ge\dimen@\dp\z@
                    \vbox to\vsize{\unvbox\z@\kern-\dimen@}%
               \else\box\z@\nobreak\bigskip
               \fi}%
     \fi
     \endgroup}


\def\cases#1{\left\{\,\vcenter{\m@th
     \ialign{$##\hfil$&\quad##\hfil\crcr#1\crcr}}\right.}
\def\matrix#1{\null\,\vcenter{\m@th
     \ialign{\hfil$##$\hfil&&\quad\hfil$##$\hfil\crcr
          \mathstrut\crcr
          \noalign{\kern-\baselineskip}
          #1\crcr
          \mathstrut\crcr
          \noalign{\kern-\baselineskip}}}\,}


\newif\ifraggedbottom

\def\raggedbottom{\ifraggedbottom\else
     \advance\topskip by\z@ plus60pt \raggedbottomtrue\fi}%
\def\normalbottom{\ifraggedbottom
     \advance\topskip by\z@ plus-60pt \raggedbottomfalse\fi}

\message{hacks,}


\toksdef\toks@i=1 \toksdef\toks@ii=2


\def\TeX{T\kern-.1667em \lower.5ex \hbox{E}\kern-.125em X\null}
\def\jyTeX{{\leavevmode
     \raise.587ex \hbox{\it\j}\kern-.1em \lower.048ex \hbox{\it y}\kern-.12em
     \TeX}}

\let\then=\iftrue
\def\ifnoarg#1\then{\def\hack@{#1}\ifx\hack@\empty}
\def\ifundefined#1\then{%
     \expandafter\ifx\csname\expandafter\blank\string#1\endcsname\relax}
\def\useif#1\then{\csname#1\endcsname}
\def\usename#1{\csname#1\endcsname}
\def\useafter#1#2{\expandafter#1\csname#2\endcsname}

\long\def\loop#1\repeat{\def\@iterate{#1\expandafter\@iterate\fi}\@iterate
     \let\@iterate=\relax}

\let\TeXend=\end
\def\begin#1{\begingroup\def\@@blockname{#1}\usename{begin#1}}
\def\end#1{\usename{end#1}\def\hack@{#1}%
     \ifx\@@blockname\hack@
          \endgroup
     \else\err@badgroup\hack@\@@blockname
     \fi}
\def\@@blockname{}

\def\defaultoption[#1]#2{%
     \def\hack@{\ifx\hack@ii[\toks@={#2}\else\toks@={#2[#1]}\fi\the\toks@}%
     \futurelet\hack@ii\hack@}

\def\markup#1{\let\@@marksf=\empty
     \ifhmode\edef\@@marksf{\spacefactor=\the\spacefactor\relax}\/\fi
     ${}^{\hbox{\subscriptfonts#1}}$\@@marksf}


\newtoks\shortyear
\newtoks\militaryhour
\newtoks\standardhour
\newtoks\minute
\newtoks\amorpm

\def\settime{\count@=\time\divide\count@ by60
     \militaryhour=\expandafter{\number\count@}%
     {\multiply\count@ by-60 \advance\count@ by\time
          \xdef\hack@{\ifnum\count@<10 0\fi\number\count@}}%
     \minute=\expandafter{\hack@}%
     \ifnum\count@<12
          \amorpm={am}
     \else\amorpm={pm}
          \ifnum\count@>12 \advance\count@ by-12 \fi
     \fi
     \standardhour=\expandafter{\number\count@}%
     \def\hack@19##1##2{\shortyear={##1##2}}%
          \expandafter\hack@\the\year}

\def\monthword#1{%
     \ifcase#1
          $\bullet$\err@badcountervalue{monthword}%
          \or January\or February\or March\or April\or May\or June%
          \or July\or August\or September\or October\or November\or December%
     \else$\bullet$\err@badcountervalue{monthword}%
     \fi}

\def\monthabbr#1{%
     \ifcase#1
          $\bullet$\err@badcountervalue{monthabbr}%
          \or Jan\or Feb\or Mar\or Apr\or May\or Jun%
          \or Jul\or Aug\or Sep\or Oct\or Nov\or Dec%
     \else$\bullet$\err@badcountervalue{monthabbr}%
     \fi}

\def\militarytime{\the\militaryhour:\the\minute}
\def\standardtime{\the\standardhour:\the\minute}


\def\@setnumstyle#1#2{\expandafter\global\expandafter\expandafter
     \expandafter\let\expandafter\expandafter
     \csname @\expandafter\blank\string#1style\endcsname
     \csname#2\endcsname}
\def\numstyle#1{\usename{@\expandafter\blank\string#1style}#1}
\def\ifblank#1\then{\useafter\ifx{@\expandafter\blank\string#1}\blank}

\def\blank#1{}

\def\Roman#1{\expandafter\uppercase\expandafter{\romannumeral#1}}
\def\alphabetic#1{%
     \ifcase#1
          $\bullet$\err@badcountervalue{alphabetic}%
          \or a\or b\or c\or d\or e\or f\or g\or h\or i\or j\or k\or l\or m%
          \or n\or o\or p\or q\or r\or s\or t\or u\or v\or w\or x\or y\or z%
     \else$\bullet$\err@badcountervalue{alphabetic}%
     \fi}
\def\Alphabetic#1{\expandafter\uppercase\expandafter{\alphabetic{#1}}}
\def\symbols#1{%
     \ifcase#1
          $\bullet$\err@badcountervalue{symbols}%
          \or*\or\dag\or\ddag\or\S\or$\|$%
          \or**\or\dag\dag\or\ddag\ddag\or\S\S\or$\|\|$%
     \else$\bullet$\err@badcountervalue{symbols}%
     \fi}


\catcode`\^^?=13 \def^^?{\relax}

\def\trimleading#1\to#2{\edef#2{#1}%
     \expandafter\@trimleading\expandafter#2#2^^?^^?}
\def\@trimleading#1#2#3^^?{\ifx#2^^?\def#1{}\else\def#1{#2#3}\fi}

\def\trimtrailing#1\to#2{\edef#2{#1}%
     \expandafter\@trimtrailing\expandafter#2#2^^? ^^?\relax}
\def\@trimtrailing#1#2 ^^?#3{\ifx#3\relax\toks@={}%
     \else\def#1{#2}\toks@={\trimtrailing#1\to#1}\fi
     \the\toks@}

\def\trim#1\to#2{\trimleading#1\to#2\trimtrailing#2\to#2}

\catcode`\^^?=15


\long\def\additemL#1\to#2{\toks@={\^^\{#1}}\toks@ii=\expandafter{#2}%
     \xdef#2{\the\toks@\the\toks@ii}}

\long\def\additemR#1\to#2{\toks@={\^^\{#1}}\toks@ii=\expandafter{#2}%
     \xdef#2{\the\toks@ii\the\toks@}}

\def\getitemL#1\to#2{\expandafter\@getitemL#1\hack@#1#2}
\def\@getitemL\^^\#1#2\hack@#3#4{\def#4{#1}\def#3{#2}}

\message{font macros,}


\newdimen\rp@
\newcount\@@sizeindex \@@sizeindex=0
\newcount\@@factori
\newcount\@@factorii
\newcount\@@factoriii
\newcount\@@factoriv

\countdef\maxfam=18
\newfam\itfam
\newfam\bffam
\newfam\bfsfam
\newfam\bmitfam

\def\@mathfontinit{\count@=4
     \loop\textfont\count@=\nullfont
          \scriptfont\count@=\nullfont
          \scriptscriptfont\count@=\nullfont
          \ifnum\count@<\maxfam\advance\count@ by\@ne
     \repeat}

\def\@fontstyleinit{%
     \def\it{\err@fontnotavailable\it}%
     \def\bf{\err@fontnotavailable\bf}%
     \def\bfs{\err@bfstobf}%
     \def\bmit{\err@fontnotavailable\bmit}%
     \def\sc{\err@fontnotavailable\sc}%
     \def\sl{\err@sltoit}%
     \def\ss{\err@fontnotavailable\ss}%
     \def\tt{\err@fontnotavailable\tt}}

\def\@parameterinit#1{\rm\rp@=.1em \@getscaling{#1}%
     \let\^^\=\@doscaling\scalingskipslist
     \setbox\strutbox=\hbox{\vrule
          height.708\baselineskip depth.292\baselineskip width\z@}}

\def\@getfactor#1#2#3#4{\@@factori=#1 \@@factorii=#2
     \@@factoriii=#3 \@@factoriv=#4}

\def\@getscaling#1{\count@=#1 \advance\count@ by-\@@sizeindex\@@sizeindex=#1
     \ifnum\count@<0
          \let\@mulordiv=\divide
          \let\@divormul=\multiply
          \multiply\count@ by\m@ne
     \else\let\@mulordiv=\multiply
          \let\@divormul=\divide
     \fi
     \edef\@@scratcha{\ifcase\count@                {1}{1}{1}{1}\or
          {1}{7}{23}{3}\or     {2}{5}{3}{1}\or      {9}{89}{13}{1}\or
          {6}{25}{6}{1}\or     {8}{71}{14}{1}\or    {6}{25}{36}{5}\or
          {1}{7}{53}{4}\or     {12}{125}{108}{5}\or {3}{14}{53}{5}\or
          {6}{41}{17}{1}\or    {13}{31}{13}{2}\or   {9}{107}{71}{2}\or
          {11}{139}{124}{3}\or {1}{6}{43}{2}\or     {10}{107}{42}{1}\or
          {1}{5}{43}{2}\or     {5}{69}{65}{1}\or    {11}{97}{91}{2}\fi}%
     \expandafter\@getfactor\@@scratcha}

\def\@doscaling#1{\@mulordiv#1by\@@factori\@divormul#1by\@@factorii
     \@mulordiv#1by\@@factoriii\@divormul#1by\@@factoriv}


\newskip\headskip
\newskip\footskip

\def\typesize=#1pt{\count@=#1 \advance\count@ by-10
     \ifcase\count@
          \@setsizex\or\err@badtypesize\or
          \@setsizexii\or\err@badtypesize\or
          \@setsizexiv
     \else\err@badtypesize
     \fi}

\def\@setsizex{\getixpt
     \def\subsubscriptfonts{\vpt}%
          \def\subsubscriptsize{\vpt\@parameterinit{-8}}%
     \def\subscriptfonts{\viipt}\def\subscriptsize{\viipt\@parameterinit{-4}}%
     \def\footnotefonts{\viiipt}\def\footnotesize{\viiipt\@parameterinit{-2}}%
     \def\smallfonts{\ixpt}\def\smallsize{\ixpt\@parameterinit{-1}}%
     \def\normalfonts{\xpt}\def\normalsize{\xpt\@parameterinit{0}}%
     \def\bigfonts{\xiipt}\def\bigsize{\xiipt\@parameterinit{2}}%
     \def\Bigfonts{\xivpt}\def\Bigsize{\xivpt\@parameterinit{4}}%
     \def\biggfonts{\xviipt}\def\biggsize{\xviipt\@parameterinit{6}}%
     \def\Biggfonts{\xxipt}\def\Biggsize{\xxipt\@parameterinit{8}}%
     \def\tinyfonts{\vpt}\def\tinysize{\vpt\@parameterinit{-8}}%
     \def\HUGEFONTS{\xxvpt}\def\HUGESIZE{\xxvpt\@parameterinit{10}}%
     \normalsize\fixedskipslist}

\def\@setsizexii{\getxipt
     \def\subsubscriptfonts{\vipt}%
          \def\subsubscriptsize{\vipt\@parameterinit{-6}}%
     \def\subscriptfonts{\viiipt}%
          \def\subscriptsize{\viiipt\@parameterinit{-2}}%
     \def\footnotefonts{\xpt}\def\footnotesize{\xpt\@parameterinit{0}}%
     \def\smallfonts{\xipt}\def\smallsize{\xipt\@parameterinit{1}}%
     \def\normalfonts{\xiipt}\def\normalsize{\xiipt\@parameterinit{2}}%
     \def\bigfonts{\xivpt}\def\bigsize{\xivpt\@parameterinit{4}}%
     \def\Bigfonts{\xviipt}\def\Bigsize{\xviipt\@parameterinit{6}}%
     \def\biggfonts{\xxipt}\def\biggsize{\xxipt\@parameterinit{8}}%
     \def\Biggfonts{\xxvpt}\def\Biggsize{\xxvpt\@parameterinit{10}}%
     \def\tinyfonts{\vpt}\def\tinysize{\vpt\@parameterinit{-8}}%
     \def\HUGEFONTS{\xxvpt}\def\HUGESIZE{\xxvpt\@parameterinit{10}}%
     \normalsize\fixedskipslist}

\def\@setsizexiv{\getxiiipt
     \def\subsubscriptfonts{\viipt}%
          \def\subsubscriptsize{\viipt\@parameterinit{-4}}%
     \def\subscriptfonts{\xpt}\def\subscriptsize{\xpt\@parameterinit{0}}%
     \def\footnotefonts{\xiipt}\def\footnotesize{\xiipt\@parameterinit{2}}%
     \def\smallfonts{\xiiipt}\def\smallsize{\xiiipt\@parameterinit{3}}%
     \def\normalfonts{\xivpt}\def\normalsize{\xivpt\@parameterinit{4}}%
     \def\bigfonts{\xviipt}\def\bigsize{\xviipt\@parameterinit{6}}%
     \def\Bigfonts{\xxipt}\def\Bigsize{\xxipt\@parameterinit{8}}%
     \def\biggfonts{\xxvpt}\def\biggsize{\xxvpt\@parameterinit{10}}%
     \def\Biggfonts{\err@sizetoolarge\Biggfonts\HUGEFONTS}%
          \def\Biggsize{\err@sizetoolarge\Biggsize\HUGESIZE}%
     \def\tinyfonts{\vpt}\def\tinysize{\vpt\@parameterinit{-8}}%
     \def\HUGEFONTS{\xxvpt}\def\HUGESIZE{\xxvpt\@parameterinit{10}}%
     \normalsize\fixedskipslist}

\def\subsubscriptfonts{\vpt} \def\subsubscriptsize{\vpt\@parameterinit{-8}}
\def\subscriptfonts{\viipt}  \def\subscriptsize{\viipt\@parameterinit{-4}}
\def\footnotefonts{\viiipt}  \def\footnotesize{\viiipt\@parameterinit{-2}}
\def\smallfonts{\err@sizenotavailable\smallfonts}
                             \def\smallsize{\ixpt\@parameterinit{-1}}
\def\normalfonts{\xpt}       \def\normalsize{\xpt\@parameterinit{0}}
\def\bigfonts{\xiipt}        \def\bigsize{\xiipt\@parameterinit{2}}
\def\Bigfonts{\xivpt}        \def\Bigsize{\xivpt\@parameterinit{4}}
\def\biggfonts{\xviipt}      \def\biggsize{\xviipt\@parameterinit{6}}
\def\Biggfonts{\xxipt}       \def\Biggsize{\xxipt\@parameterinit{8}}
\def\tinyfonts{\vpt}         \def\tinysize{\vpt\@parameterinit{-8}}
\def\HUGEFONTS{\xxvpt}       \def\HUGESIZE{\xxvpt\@parameterinit{10}}

\message{document layout,}


\newtoks\everyoutput \everyoutput={}
\newdimen\depthofpage
\newcount\pagenum \pagenum=0

\newdimen\oddtopmargin  \newdimen\eventopmargin
\newdimen\oddleftmargin \newdimen\evenleftmargin
\newtoks\oddhead        \newtoks\evenhead
\newtoks\oddfoot        \newtoks\evenfoot

\def\topmargin{\afterassignment\@seteventop\oddtopmargin}
\def\leftmargin{\afterassignment\@setevenleft\oddleftmargin}
\def\head{\afterassignment\@setevenhead\oddhead}
\def\foot{\afterassignment\@setevenfoot\oddfoot}

\def\@seteventop{\eventopmargin=\oddtopmargin}
\def\@setevenleft{\evenleftmargin=\oddleftmargin}
\def\@setevenhead{\evenhead=\oddhead}
\def\@setevenfoot{\evenfoot=\oddfoot}

\def\pagenumstyle#1{\@setnumstyle\pagenum{#1}}

\newif\ifdraft
\def\draft{\drafttrue\leftmargin=.5in \overfullrule=5pt }

\def\outputstyle#1{\global\expandafter\let\expandafter
          \@outputstyle\csname#1output\endcsname
     \usename{#1setup}}

\output={\@outputstyle}

\def\normaloutput{\the\everyoutput
     \global\advance\pagenum by\@ne
     \ifodd\pagenum
          \voffset=\oddtopmargin \hoffset=\oddleftmargin
     \else\voffset=\eventopmargin \hoffset=\evenleftmargin
     \fi
     \advance\voffset by-1in  \advance\hoffset by-1in
     \count0=\pagenum
     \expandafter\shipout\pagebox
     \ifnum\outputpenalty>-\@MM\else\dosupereject\fi}

\newdimen\fullhsize
\newbox\leftpage
\newcount\leftpagenum
\newcount\outputpagenum \outputpagenum=0
\let\leftorright=L

\def\twoupoutput{\the\everyoutput
     \global\advance\pagenum by\@ne
     \if L\leftorright
          \global\setbox\leftpage=\leftline{\pagebox}%
          \global\leftpagenum=\pagenum
          \global\let\leftorright=R%
     \else\global\advance\outputpagenum by\@ne
          \ifodd\outputpagenum
               \voffset=\oddtopmargin \hoffset=\oddleftmargin
          \else\voffset=\eventopmargin \hoffset=\evenleftmargin
          \fi
          \advance\voffset by-1in  \advance\hoffset by-1in
          \count0=\leftpagenum \count1=\pagenum
          \shipout\vbox{\hbox to\fullhsize
               {\box\leftpage\hfil\leftline{\pagebox}}}%
          \global\let\leftorright=L%
     \fi
     \ifnum\outputpenalty>-\@MM
     \else\dosupereject
          \if R\leftorright
               \globaldefs=\@ne\head={\hfil}\foot={\hfil}\globaldefs=\z@
               \null\newpage
          \fi
     \fi}

\def\pagebox{\vbox{\makeheadline\pagebody\makefootline}}

\def\makeheadline{%
     \vbox to\z@{\baselinestretch=\@m
          \vskip\topskip\vskip-.708\baselineskip\vskip-\headskip
          \line{\vbox to\ht\strutbox{}%
               \ifodd\pagenum\the\oddhead\else\the\evenhead\fi}%
          \vss}%
     \nointerlineskip}

\def\pagebody{\vbox to\vsize{%
     \boxmaxdepth\maxdepth
     \ifvoid\topins\else\unvbox\topins\fi
     \depthofpage=\dp255
     \unvbox255
     \ifraggedbottom\kern-\depthofpage\vfil\fi
     \ifvoid\footins
     \else\vskip\skip\footins
          \footnoterule
          \unvbox\footins
          \vskip-\footnoteskip
     \fi}}

\def\makefootline{\baselineskip=\footskip
     \line{\ifodd\pagenum\the\oddfoot\else\the\evenfoot\fi}}


\newskip\abovechapterskip
\newskip\belowchapterskip
\newskip\abovesectionskip
\newskip\belowsectionskip
\newskip\abovesubsectionskip
\newskip\belowsubsectionskip

\def\chapterstyle#1{\global\expandafter\let\expandafter\@chapterstyle
     \csname#1text\endcsname}
\def\sectionstyle#1{\global\expandafter\let\expandafter\@sectionstyle
     \csname#1text\endcsname}
\def\subsectionstyle#1{\global\expandafter\let\expandafter\@subsectionstyle
     \csname#1text\endcsname}

\def\chapter#1{%
     \ifdim\lastskip=17sp \else\chapterbreak\vskip\abovechapterskip\fi
     \@chapterstyle{\ifblank\chapternumstyle\then
          \else\newchapternum=\next\chapternumformat\ \fi#1}%
     \nobreak\vskip\belowchapterskip\vskip17sp }

\def\section#1{%
     \ifdim\lastskip=17sp \else\sectionbreak\vskip\abovesectionskip\fi
     \@sectionstyle{\ifblank\sectionnumstyle\then
          \else\newsectionnum=\next\sectionnumformat\ \fi#1}%
     \nobreak\vskip\belowsectionskip\vskip17sp }

\def\subsection#1{%
     \ifdim\lastskip=17sp \else\subsectionbreak\vskip\abovesubsectionskip\fi
     \@subsectionstyle{\ifblank\subsectionnumstyle\then
          \else\newsubsectionnum=\next\subsectionnumformat\ \fi#1}%
     \nobreak\vskip\belowsubsectionskip\vskip17sp }


\let\TeXunderline=\underline
\let\TeXoverline=\overline
\def\underline#1{\relax\ifmmode\TeXunderline{#1}\else
     $\TeXunderline{\hbox{#1}}$\fi}
\def\overline#1{\relax\ifmmode\TeXoverline{#1}\else
     $\TeXoverline{\hbox{#1}}$\fi}

\def\baselinestretch{\afterassignment\@baselinestretch\count@}
\def\@baselinestretch{\baselineskip=\normalbaselineskip
     \divide\baselineskip by\@m\baselineskip=\count@\baselineskip
     \setbox\strutbox=\hbox{\vrule
          height.708\baselineskip depth.292\baselineskip width\z@}%
     \bigskipamount=\the\baselineskip
          plus.25\baselineskip minus.25\baselineskip
     \medskipamount=.5\baselineskip
          plus.125\baselineskip minus.125\baselineskip
     \smallskipamount=.25\baselineskip
          plus.0625\baselineskip minus.0625\baselineskip}

\def\\{\ifhmode\ifnum\lastpenalty=-\@M\else\hfil\penalty-\@M\fi\fi
     \ignorespaces}
\def\newpage{\vfil\break}

\def\lefttext#1{\par{\@text\leftskip=\z@\rightskip=\centering
     \noindent#1\par}}
\def\righttext#1{\par{\@text\leftskip=\centering\rightskip=\z@
     \noindent#1\par}}
\def\centertext#1{\par{\@text\leftskip=\centering\rightskip=\centering
     \noindent#1\par}}
\def\@text{\parindent=\z@ \parfillskip=\z@ \everypar={}%
     \spaceskip=.3333em \xspaceskip=.5em
     \def\\{\ifhmode\ifnum\lastpenalty=-\@M\else\penalty-\@M\fi\fi
          \ignorespaces}}

\def\beginleft{\par\@text\leftskip=\z@ \rightskip=\centering}
     
\def\beginright{\par\@text\leftskip=\centering\rightskip=\z@ }
     
\def\begincenter{\par\@text\leftskip=\centering\rightskip=\centering}

\def\beginnarrow{\defaultoption[\parindent]\@beginnarrow}
\def\@beginnarrow[#1]{\par\advance\leftskip by#1\advance\rightskip by#1}

\begingroup
\catcode`\[=1 \catcode`\{=11 \gdef\beginignore[\endgroup\bgroup
     \catcode`\e=0 \catcode`\\=12 \catcode`\{=11 \catcode`\f=12 \let\or=\relax
     \let\nd{ignor=\fi \let\}=\egroup
     \iffalse}
\endgroup

\long\def\marginnote#1{\leavevmode
     \edef\@marginsf{\spacefactor=\the\spacefactor\relax}%
     \ifdraft\strut\vadjust{%
          \hbox to\z@{\hskip\hsize\hskip.1in
               \vbox to\z@{\vskip-\dp\strutbox
                    \marginnoteformat
                    \vskip-\ht\strutbox
                    \noindent\strut#1\par
                    \vss}%
               \hss}}%
     \fi
     \@marginsf}


\newtoks\everybye \everybye={\par\vfil}
\outer\def\bye{\the\everybye
     \footnotecheck
     \prelabelcheck
     \streamcheck
     \supereject
     \TeXend}

\message{footnotes,}

\newcount\footnotenum \footnotenum=0
\newskip\footnoteskip
\let\@footnotelist=\empty

\def\footnotenumstyle#1{\@setnumstyle\footnotenum{#1}%
     \useafter\ifx{@footnotenumstyle}\symbols
          \global\let\@footup=\empty
     \else\global\let\@footup=\markup
     \fi}

\def\footnote{\footnotecheck\defaultoption[]\@footnote}
\def\@footnote[#1]{\@footnotemark[#1]\@footnotetext}

\def\footnotemark{\defaultoption[]\@footnotemark}
\def\@footnotemark[#1]{\let\@footsf=\empty
     \ifhmode\edef\@footsf{\spacefactor=\the\spacefactor\relax}\/\fi
     \ifnoarg#1\then
          \global\advance\footnotenum by\@ne
          \@footup{\footnotenumformat}%
          \edef\@@foota{\footnotenum=\the\footnotenum\relax}%
          \expandafter\additemR\expandafter\@footup\expandafter
               {\@@foota\footnotenumformat}\to\@footnotelist
          \global\let\@footnotelist=\@footnotelist
     \else\markup{#1}%
          \additemR\markup{#1}\to\@footnotelist
          \global\let\@footnotelist=\@footnotelist
     \fi
     \@footsf}

\def\footnotetext{%
     \ifx\@footnotelist\empty\err@extrafootnotetext\else\@footnotetext\fi}
\def\@footnotetext{%
     \getitemL\@footnotelist\to\@@foota
     \global\let\@footnotelist=\@footnotelist
     \insert\footins\bgroup
     \footnoteformat
     \splittopskip=\ht\strutbox\splitmaxdepth=\dp\strutbox
     \interlinepenalty=\interfootnotelinepenalty\floatingpenalty=\@MM
     \noindent\llap{\@@foota}\strut
     \bgroup\aftergroup\@footnoteend
     \let\@@scratcha=}
\def\@footnoteend{\strut\par\vskip\footnoteskip\egroup}

\def\footnoterule{\normalfonts
     \kern-.3em \hrule width2in height.04em \kern .26em }

\def\footnotecheck{%
     \ifx\@footnotelist\empty
     \else\err@extrafootnotemark
          \global\let\@footnotelist=\empty
     \fi}

\message{labels,}

\let\@@labeldef=\xdef
\newif\if@labelfile
\newwrite\@labelfile
\let\@prelabellist=\empty

\def\label#1#2{\trim#1\to\@@labarg\edef\@@labtext{#2}%
     \edef\@@labname{lab@\@@labarg}%
     \useafter\ifundefined\@@labname\then\else\@yeslab\fi
     \useafter\@@labeldef\@@labname{#2}%
     \ifstreaming
          \expandafter\toks@\expandafter\expandafter\expandafter
               {\csname\@@labname\endcsname}%
          \immediate\write\streamout{\noexpand\label{\@@labarg}{\the\toks@}}%
     \fi}
\def\@yeslab{%
     \useafter\ifundefined{if\@@labname}\then
          \err@labelredef\@@labarg
     \else\useif{if\@@labname}\then
               \err@labelredef\@@labarg
          \else\global\usename{\@@labname true}%
               \useafter\ifundefined{pre\@@labname}\then
               \else\useafter\ifx{pre\@@labname}\@@labtext
                    \else\err@badlabelmatch\@@labarg
                    \fi
               \fi
               \if@labelfile
               \else\global\@labelfiletrue
                    \immediate\write\sixt@@n{--> Creating file \jobname.lab}%
                    \immediate\openout\@labelfile=\jobname.lab
               \fi
               \immediate\write\@labelfile
                    {\noexpand\prelabel{\@@labarg}{\@@labtext}}%
          \fi
     \fi}

\def\putlab#1{\trim#1\to\@@labarg\edef\@@labname{lab@\@@labarg}%
     \useafter\ifundefined\@@labname\then\@nolab\else\usename\@@labname\fi}
\def\@nolab{%
     \useafter\ifundefined{pre\@@labname}\then
          \undefinedlabelformat
          \err@needlabel\@@labarg
          \useafter\xdef\@@labname{\undefinedlabelformat}%
     \else\usename{pre\@@labname}%
          \useafter\xdef\@@labname{\usename{pre\@@labname}}%
     \fi
     \useafter\newif{if\@@labname}%
     \expandafter\additemR\@@labarg\to\@prelabellist}

\def\prelabel#1{\useafter\gdef{prelab@#1}}

\def\ifundefinedlabel#1\then{%
     \expandafter\ifx\csname lab@#1\endcsname\relax}
\def\useiflab#1\then{\csname iflab@#1\endcsname}

\def\prelabelcheck{{%
     \def\^^\##1{\useiflab{##1}\then\else\err@undefinedlabel{##1}\fi}%
     \@prelabellist}}

\message{equation numbering,}

\newcount\chapternum
\newcount\sectionnum
\newcount\subsectionnum
\newcount\equationnum
\newcount\subequationnum
\newcount\figurenum
\newcount\subfigurenum
\newcount\tablenum
\newcount\subtablenum

\newif\if@subeqncount
\newif\if@subfigcount
\newif\if@subtblcount

\def\newchapternum{\newsectionnum=\z@\@resetnum\chapternum}
\def\newsectionnum{\newsubsectionnum=\z@\@resetnum\sectionnum}
\def\newsubsectionnum{\newequationnum=\z@\newfigurenum=\z@\newtablenum=\z@
     \@resetnum\subsectionnum}
\def\newequationnum{\newsubequationnum=\z@\@resetnum\equationnum}
\def\newsubequationnum{\@resetnum\subequationnum}
\def\newfigurenum{\newsubfigurenum=\z@\@resetnum\figurenum}
\def\newsubfigurenum{\@resetnum\subfigurenum}
\def\newtablenum{\newsubtablenum=\z@\@resetnum\tablenum}
\def\newsubtablenum{\@resetnum\subtablenum}

\def\@resetnum#1{\global\advance#1by1 \edef\next{\the#1\relax}\global#1}

\newchapternum=0

\def\chapternumstyle#1{\@setnumstyle\chapternum{#1}}
\def\sectionnumstyle#1{\@setnumstyle\sectionnum{#1}}
\def\subsectionnumstyle#1{\@setnumstyle\subsectionnum{#1}}
\def\equationnumstyle#1{\@setnumstyle\equationnum{#1}}
\def\subequationnumstyle#1{\@setnumstyle\subequationnum{#1}%
     \ifblank\subequationnumstyle\then\global\@subeqncountfalse\fi
     \ignorespaces}
\def\figurenumstyle#1{\@setnumstyle\figurenum{#1}}
\def\subfigurenumstyle#1{\@setnumstyle\subfigurenum{#1}%
     \ifblank\subfigurenumstyle\then\global\@subfigcountfalse\fi
     \ignorespaces}
\def\tablenumstyle#1{\@setnumstyle\tablenum{#1}}
\def\subtablenumstyle#1{\@setnumstyle\subtablenum{#1}%
     \ifblank\subtablenumstyle\then\global\@subtblcountfalse\fi
     \ignorespaces}

\def\eqnlabel#1{%
     \if@subeqncount
          \newsubequationnum=\next
     \else\newequationnum=\next
          \ifblank\subequationnumstyle\then
          \else\global\@subeqncounttrue
               \newsubequationnum=\@ne
          \fi
     \fi
     \label{#1}{\puteqnformat}(\puteqn{#1})%
     \ifdraft\rlap{\hskip.1in{\tt#1}}\fi}

\let\puteqn=\putlab

\def\equation#1#2{\useafter\gdef{eqn@#1}{#2\eqno\eqnlabel{#1}}}
\def\Equation#1{\useafter\gdef{eqn@#1}}

\def\putequation#1{\useafter\ifundefined{eqn@#1}\then
     \err@undefinedeqn{#1}\else\usename{eqn@#1}\fi}

\def\eqnseriesstyle#1{\gdef\@eqnseriesstyle{#1}}
\def\begineqnseries{\subequationnumstyle{\@eqnseriesstyle}%
     \defaultoption[]\@begineqnseries}
\def\@begineqnseries[#1]{\edef\@@eqnname{#1}}
\def\endeqnseries{\subequationnumstyle{blank}%
     \expandafter\ifnoarg\@@eqnname\then
     \else\label\@@eqnname{\puteqnformat}%
     \fi
     \aftergroup\ignorespaces}

\def\figlabel#1{%
     \if@subfigcount
          \newsubfigurenum=\next
     \else\newfigurenum=\next
          \ifblank\subfigurenumstyle\then
          \else\global\@subfigcounttrue
               \newsubfigurenum=\@ne
          \fi
     \fi
     \label{#1}{\putfigformat}\putfig{#1}%
     {\def\marginnoteformat{\tt}\marginnote{#1}}}

\let\putfig=\putlab

\def\figseriesstyle#1{\gdef\@figseriesstyle{#1}}
\def\beginfigseries{\subfigurenumstyle{\@figseriesstyle}%
     \defaultoption[]\@beginfigseries}
\def\@beginfigseries[#1]{\edef\@@figname{#1}}
\def\endfigseries{\subfigurenumstyle{blank}%
     \expandafter\ifnoarg\@@figname\then
     \else\label\@@figname{\putfigformat}%
     \fi
     \aftergroup\ignorespaces}

\def\tbllabel#1{%
     \if@subtblcount
          \newsubtablenum=\next
     \else\newtablenum=\next
          \ifblank\subtablenumstyle\then
          \else\global\@subtblcounttrue
               \newsubtablenum=\@ne
          \fi
     \fi
     \label{#1}{\puttblformat}\puttbl{#1}%
     {\def\marginnoteformat{\tt}\marginnote{#1}}}

\let\puttbl=\putlab

\def\tblseriesstyle#1{\gdef\@tblseriesstyle{#1}}
\def\begintblseries{\subtablenumstyle{\@tblseriesstyle}%
     \defaultoption[]\@begintblseries}
\def\@begintblseries[#1]{\edef\@@tblname{#1}}
\def\endtblseries{\subtablenumstyle{blank}%
     \expandafter\ifnoarg\@@tblname\then
     \else\label\@@tblname{\puttblformat}%
     \fi
     \aftergroup\ignorespaces}

\message{reference numbering,}

\newcount\referencenum \referencenum=0
\newcount\@@prerefcount \@@prerefcount=0
\newcount\@@thisref
\newcount\@@lastref
\newcount\@@loopref
\newcount\@@refseq
\newdimen\refnumindent
\let\@undefreflist=\empty

\def\referencenumstyle#1{\@setnumstyle\referencenum{#1}}

\def\referencestyle#1{\usename{@ref#1}}

\def\@refsequential{%
     \gdef\@refpredef##1{\global\advance\referencenum by\@ne
          \let\^^\=0\label{##1}{\^^\{\the\referencenum}}%
          \useafter\gdef{ref@\the\referencenum}{{##1}{\undefinedlabelformat}}}%
     \gdef\@reference##1##2{%
          \ifundefinedlabel##1\then
          \else\def\^^\####1{\global\@@thisref=####1\relax}\putlab{##1}%
               \useafter\gdef{ref@\the\@@thisref}{{##1}{##2}}%
          \fi}%
     \gdef\endputreferences{%
          \loop\ifnum\@@loopref<\referencenum
                    \advance\@@loopref by\@ne
                    \expandafter\expandafter\expandafter\@printreference
                         \csname ref@\the\@@loopref\endcsname
          \repeat
          \par}}

\def\@refpreordered{%
     \gdef\@refpredef##1{\global\advance\referencenum by\@ne
          \additemR##1\to\@undefreflist}%
     \gdef\@reference##1##2{%
          \ifundefinedlabel##1\then
          \else\global\advance\@@loopref by\@ne
               {\let\^^\=0\label{##1}{\^^\{\the\@@loopref}}}%
               \@printreference{##1}{##2}%
          \fi}
     \gdef\endputreferences{%
          \def\^^\####1{\useiflab{####1}\then
               \else\reference{####1}{\undefinedlabelformat}\fi}%
          \@undefreflist
          \par}}

\def\beginprereferences{\par
     \def\reference##1##2{\global\advance\referencenum by1\@ne
          \let\^^\=0\label{##1}{\^^\{\the\referencenum}}%
          \useafter\gdef{ref@\the\referencenum}{{##1}{##2}}}}
\def\endprereferences{\global\@@prerefcount=\the\referencenum\par}

\def\beginputreferences{\par
     \refnumindent=\z@\@@loopref=\z@
     \loop\ifnum\@@loopref<\referencenum
               \advance\@@loopref by\@ne
               \setbox\z@=\hbox{\referencenum=\@@loopref
                    \referencenumformat\enskip}%
               \ifdim\wd\z@>\refnumindent\refnumindent=\wd\z@\fi
     \repeat
     \putreferenceformat
     \@@loopref=\z@
     \loop\ifnum\@@loopref<\@@prerefcount
               \advance\@@loopref by\@ne
               \expandafter\expandafter\expandafter\@printreference
                    \csname ref@\the\@@loopref\endcsname
     \repeat
     \let\reference=\@reference}

\def\@printreference#1#2{\ifx#2\undefinedlabelformat\err@undefinedref{#1}\fi
     \noindent\ifdraft\rlap{\hskip\hsize\hskip.1in \tt#1}\fi
     \llap{\referencenum=\@@loopref\referencenumformat\enskip}#2\par}

\def\reference#1#2{{\par\refnumindent=\z@\putreferenceformat\noindent#2\par}}

\def\putref#1{\trim#1\to\@@refarg
     \expandafter\ifnoarg\@@refarg\then
          \toks@={\relax}%
     \else\@@lastref=-\@m\def\@@refsep{}\def\@more{\@nextref}%
          \toks@={\@nextref#1,,}%
     \fi\the\toks@}
\def\@nextref#1,{\trim#1\to\@@refarg
     \expandafter\ifnoarg\@@refarg\then
          \let\@more=\relax
     \else\ifundefinedlabel\@@refarg\then
               \expandafter\@refpredef\expandafter{\@@refarg}%
          \fi
          \def\^^\##1{\global\@@thisref=##1\relax}%
          \global\@@thisref=\m@ne
          \setbox\z@=\hbox{\putlab\@@refarg}%
     \fi
     \advance\@@lastref by\@ne
     \ifnum\@@lastref=\@@thisref\advance\@@refseq by\@ne\else\@@refseq=\@ne\fi
     \ifnum\@@lastref<\z@
     \else\ifnum\@@refseq<\thr@@
               \@@refsep\def\@@refsep{,}%
               \ifnum\@@lastref>\z@
                    \advance\@@lastref by\m@ne
                    {\referencenum=\@@lastref\putrefformat}%
               \else\undefinedlabelformat
               \fi
          \else\def\@@refsep{--}%
          \fi
     \fi
     \@@lastref=\@@thisref
     \@more}

\message{streaming,}

\newif\ifstreaming

\def\streamto{\defaultoption[\jobname]\@streamto}
\def\@streamto[#1]{\global\streamingtrue
     \immediate\write\sixt@@n{--> Streaming to #1.str}%
     \newwrite\streamout\immediate\openout\streamout=#1.str }

\def\streamfrom{\defaultoption[\jobname]\@streamfrom}
\def\@streamfrom[#1]{\newread\streamin\openin\streamin=#1.str
     \ifeof\streamin
          \expandafter\err@nostream\expandafter{#1.str}%
     \else\immediate\write\sixt@@n{--> Streaming from #1.str}%
          \let\@@labeldef=\gdef
          \ifstreaming
               \edef\@elc{\endlinechar=\the\endlinechar}%
               \endlinechar=\m@ne
               \loop\read\streamin to\@@scratcha
                    \ifeof\streamin
                         \streamingfalse
                    \else\toks@=\expandafter{\@@scratcha}%
                         \immediate\write\streamout{\the\toks@}%
                    \fi
                    \ifstreaming
               \repeat
               \@elc
               \input #1.str
               \streamingtrue
          \else\input #1.str
          \fi
          \let\@@labeldef=\xdef
     \fi}

\def\streamcheck{\ifstreaming
     \immediate\write\streamout{\pagenum=\the\pagenum}%
     \immediate\write\streamout{\footnotenum=\the\footnotenum}%
     \immediate\write\streamout{\referencenum=\the\referencenum}%
     \immediate\write\streamout{\chapternum=\the\chapternum}%
     \immediate\write\streamout{\sectionnum=\the\sectionnum}%
     \immediate\write\streamout{\subsectionnum=\the\subsectionnum}%
     \immediate\write\streamout{\equationnum=\the\equationnum}%
     \immediate\write\streamout{\subequationnum=\the\subequationnum}%
     \immediate\write\streamout{\figurenum=\the\figurenum}%
     \immediate\write\streamout{\subfigurenum=\the\subfigurenum}%
     \immediate\write\streamout{\tablenum=\the\tablenum}%
     \immediate\write\streamout{\subtablenum=\the\subtablenum}%
     \immediate\closeout\streamout
     \fi}


\def\err@badtypesize{%
     \errhelp={The limited availability of certain fonts requires^^J%
          that the base type size be 10pt, 12pt, or 14pt.^^J}%
     \errmessage{--> Illegal base type size}}

\def\err@badsizechange{\immediate\write\sixt@@n
     {--> Size change not allowed in math mode, ignored}}

\def\err@sizetoolarge#1{\immediate\write\sixt@@n
     {--> \noexpand#1 too big, substituting HUGE}}

\def\err@sizenotavailable#1{\immediate\write\sixt@@n
     {--> Size not available, \noexpand#1 ignored}}

\def\err@fontnotavailable#1{\immediate\write\sixt@@n
     {--> Font not available, \noexpand#1 ignored}}

\def\err@sltoit{\immediate\write\sixt@@n
     {--> Style \noexpand\sl not available, substituting \noexpand\it}%
     \it}

\def\err@bfstobf{\immediate\write\sixt@@n
     {--> Style \noexpand\bfs not available, substituting \noexpand\bf}%
     \bf}

\def\err@badgroup#1#2{%
     \errhelp={The block you have just tried to close was not the one^^J%
          most recently opened.^^J}%
     \errmessage{--> \noexpand\end{#1} doesn't match \noexpand\begin{#2}}}

\def\err@badcountervalue#1{\immediate\write\sixt@@n
     {--> Counter (#1) out of bounds}}

\def\err@extrafootnotemark{\immediate\write\sixt@@n
     {--> \noexpand\footnotemark command
          has no corresponding \noexpand\footnotetext}}

\def\err@extrafootnotetext{%
     \errhelp{You have given a \noexpand\footnotetext command without first
          specifying^^Ja \noexpand\footnotemark.^^J}%
     \errmessage{--> \noexpand\footnotetext command has no corresponding
          \noexpand\footnotemark}}

\def\err@labelredef#1{\immediate\write\sixt@@n
     {--> Label "#1" redefined}}

\def\err@badlabelmatch#1{\immediate\write\sixt@@n
     {--> Definition of label "#1" doesn't match value in \jobname.lab}}

\def\err@needlabel#1{\immediate\write\sixt@@n
     {--> Label "#1" cited before its definition}}

\def\err@undefinedlabel#1{\immediate\write\sixt@@n
     {--> Label "#1" cited but never defined}}

\def\err@undefinedeqn#1{\immediate\write\sixt@@n
     {--> Equation "#1" not defined}}

\def\err@undefinedref#1{\immediate\write\sixt@@n
     {--> Reference "#1" not defined}}

\def\err@nostream#1{%
     \errhelp={You have tried to input a stream file that doesn't exist.^^J}%
     \errmessage{--> Stream file #1 not found}}

\message{jyTeX initialization}

\everyjob{\immediate\write16{--> jyTeX version \fmtversion}%
     \edef\@@jobname{\jobname}%
     \edef\jobname{\@@jobname}%
     \settime
     \openin0=\jobname.lab
     \ifeof0
     \else\closein0
          \immediate\write16{--> Getting labels from file \jobname.lab}%
          \input\jobname.lab
     \fi}


\def\fixedskipslist{%
     \^^\{\topskip}%
     \^^\{\splittopskip}%
     \^^\{\maxdepth}%
     \^^\{\skip\topins}%
     \^^\{\skip\footins}%
     \^^\{\headskip}%
     \^^\{\footskip}}

\def\scalingskipslist{%
     \^^\{\p@renwd}%
     \^^\{\delimitershortfall}%
     \^^\{\nulldelimiterspace}%
     \^^\{\scriptspace}%
     \^^\{\jot}%
     \^^\{\normalbaselineskip}%
     \^^\{\normallineskip}%
     \^^\{\normallineskiplimit}%
     \^^\{\baselineskip}%
     \^^\{\lineskip}%
     \^^\{\lineskiplimit}%
     \^^\{\bigskipamount}%
     \^^\{\medskipamount}%
     \^^\{\smallskipamount}%
     \^^\{\parskip}%
     \^^\{\parindent}%
     \^^\{\abovedisplayskip}%
     \^^\{\belowdisplayskip}%
     \^^\{\abovedisplayshortskip}%
     \^^\{\belowdisplayshortskip}%
     \^^\{\abovechapterskip}%
     \^^\{\belowchapterskip}%
     \^^\{\abovesectionskip}%
     \^^\{\belowsectionskip}%
     \^^\{\abovesubsectionskip}%
     \^^\{\belowsubsectionskip}}


\def\twoupsetup{
     \topmargin=.75in
     \leftmargin=.5in
     \vsize=6.9in
     \hsize=4.75in
     \fullhsize=10in
     \let\draft=\relax}

\outputstyle{normal}                             

\def\marginnoteformat{\subscriptsize             
     \hsize=1in \baselinestretch=1000 \everypar={}%
     \tolerance=5000 \hbadness=5000 \parskip=0pt \parindent=0pt
     \leftskip=0pt \rightskip=0pt \raggedright}

\head={\ifdraft\normalfonts\it\hfil DRAFT\hfil   
     \llap{\number\day\ \monthword\month\ \militarytime}\else\hfil\fi}
\foot={\hfil\normalfonts\numstyle\pagenum\hfil}  

\normalbaselineskip=12pt                         
\normallineskip=0pt                              
\normallineskiplimit=0pt                         
\normalbaselines                                 

\topskip=.85\baselineskip \splittopskip=\topskip \headskip=2\baselineskip
\footskip=\headskip

\pagenumstyle{arabic}                            

\parskip=0pt                                     
\parindent=20pt                                  

\baselinestretch=1000                            


\chapterstyle{left}                              
\chapternumstyle{blank}                          
\def\chapterbreak{\newpage}                      
\abovechapterskip=0pt                            
\belowchapterskip=1.5\baselineskip               
     plus.38\baselineskip minus.38\baselineskip
\def\chapternumformat{\numstyle\chapternum.}     

\sectionstyle{left}                              
\sectionnumstyle{blank}                          
\def\sectionbreak{\vskip0pt plus4\baselineskip\penalty-100
     \vskip0pt plus-4\baselineskip}              
\abovesectionskip=1.5\baselineskip               
     plus.38\baselineskip minus.38\baselineskip
\belowsectionskip=\the\baselineskip              
     plus.25\baselineskip minus.25\baselineskip
\def\sectionnumformat{
     \ifblank\chapternumstyle\then\else\numstyle\chapternum.\fi
     \numstyle\sectionnum.}

\subsectionstyle{left}                           
\subsectionnumstyle{blank}                       
\def\subsectionbreak{\vskip0pt plus4\baselineskip\penalty-100
     \vskip0pt plus-4\baselineskip}              
\abovesubsectionskip=\the\baselineskip           
     plus.25\baselineskip minus.25\baselineskip
\belowsubsectionskip=.75\baselineskip            
     plus.19\baselineskip minus.19\baselineskip
\def\subsectionnumformat{
     \ifblank\chapternumstyle\then\else\numstyle\chapternum.\fi
     \ifblank\sectionnumstyle\then\else\numstyle\sectionnum.\fi
     \numstyle\subsectionnum.}


\footnotenumstyle{symbols}                       
\footnoteskip=0pt                                
\def\footnotenumformat{\numstyle\footnotenum}    
\def\footnoteformat{\footnotesize                
     \everypar={}\parskip=0pt \parfillskip=0pt plus1fil
     \leftskip=1em \rightskip=0pt
     \spaceskip=0pt \xspaceskip=0pt
     \def\\{\ifhmode\ifnum\lastpenalty=-10000
          \else\hfil\penalty-10000 \fi\fi\ignorespaces}}


\def\undefinedlabelformat{$\bullet$}             


\equationnumstyle{arabic}                        
\subequationnumstyle{blank}                      
\figurenumstyle{arabic}                          
\subfigurenumstyle{blank}                        
\tablenumstyle{arabic}                           
\subtablenumstyle{blank}                         

\eqnseriesstyle{alphabetic}                      
\figseriesstyle{alphabetic}                      
\tblseriesstyle{alphabetic}                      

\def\puteqnformat{\hbox{
     \ifblank\chapternumstyle\then\else\numstyle\chapternum.\fi
     \ifblank\sectionnumstyle\then\else\numstyle\sectionnum.\fi
     \ifblank\subsectionnumstyle\then\else\numstyle\subsectionnum.\fi
     \numstyle\equationnum
     \numstyle\subequationnum}}
\def\putfigformat{\hbox{
     \ifblank\chapternumstyle\then\else\numstyle\chapternum.\fi
     \ifblank\sectionnumstyle\then\else\numstyle\sectionnum.\fi
     \ifblank\subsectionnumstyle\then\else\numstyle\subsectionnum.\fi
     \numstyle\figurenum
     \numstyle\subfigurenum}}
\def\puttblformat{\hbox{
     \ifblank\chapternumstyle\then\else\numstyle\chapternum.\fi
     \ifblank\sectionnumstyle\then\else\numstyle\sectionnum.\fi
     \ifblank\subsectionnumstyle\then\else\numstyle\subsectionnum.\fi
     \numstyle\tablenum
     \numstyle\subtablenum}}


\referencestyle{sequential}                      
\referencenumstyle{arabic}                       
\def\putrefformat{\numstyle\referencenum}        
\def\referencenumformat{\numstyle\referencenum.} 
\def\putreferenceformat{
     \everypar={\hangindent=1em \hangafter=1 }%
     \def\\{\hfil\break\null\hskip-1em \ignorespaces}%
     \leftskip=\refnumindent\parindent=0pt \interlinepenalty=1000 }


\normalsize


\def\fmtversion{2.6M (June 1992)}

\catcode`\@=12

\typesize=10pt \magnification=1200 \baselineskip17truept
\footnotenumstyle{arabic} \hsize=6truein\vsize=8.5truein
\input epsf
\sectionnumstyle{blank}
\chapternumstyle{blank}
\chapternum=1
\sectionnum=1
\pagenum=0

\def\begintitle{\pagenumstyle{blank}\parindent=0pt
\begin{narrow}[0.4in]}
\def\endtitle{\end{narrow}\newpage\pagenumstyle{arabic}}


\def\beginexercise{\vskip 20truept\parindent=0pt\begin{narrow}[10
truept]}
\def\endexercise{\vskip 10truept\end{narrow}}


\def\eql#1{\eqno\eqnlabel{#1}}
\def\ref{\reference}
\def\peq{\puteqn}
\def\pref{\putref}

\def\mgn{\marginnote}
\def\bex{\begin{exercise}}
\def\eex{\end{exercise}}


\font\open=msbm10 

\font\ssb=cmss10

\def\StretchRtArr#1{{\count255=0\loop\relbar\joinrel\advance\count255 by1
\ifnum\count255<#1\repeat\rightarrow}}
\def\StretchLtArr#1{\,{\leftarrow\!\!\count255=0\loop\relbar
\joinrel\advance\count255 by1\ifnum\count255<#1\repeat}}

\def\StretchLRtArr#1{\,{\leftarrow\!\!\count255=0\loop\relbar\joinrel\advance
\count255 by1\ifnum\count255<#1\repeat\rightarrow\,\,}}

\def\mbox#1{{\leavevmode\hbox{#1}}}

\def\hspace#1{{\phantom{\mbox#1}}}
\def\oR{\mbox{\open\char82}}

\def\oZ{\mbox{\open\char90}}

\def\sssb{\mbox{{\ssb\char98}}}

\def\bom{{\bmit\omega}}
\def\be{\beta}

\def\Ga{\Gamma}

\def\ep{\epsilon}

\def\S{\$}

\def\la{\lambda}

\def\om{\omega}

\def\si{\sigma}

\def\ze{\zeta}

\def\De{\Delta}

\def\caB{{\cal B}}

\def\caO{{\cal O}}

\def\det{{\rm det\,}}

\def\Real{{\rm Re\,}}

\def\sc{{\rm sc }}

\def\zf{$\zeta$--function}


\def\frac#1/#2{\leavevmode\kern.1em
\raise.5ex\hbox{\the\scriptfont0 #1}\kern-.1em/\kern-.15em
\lower.25ex\hbox{\the\scriptfont0 #2}}
\def\sfrac#1/#2{\leavevmode\kern.1em
\raise.5ex\hbox{\the\scriptscriptfont0 #1}\kern-.1em/\kern-.15em
\lower.25ex\hbox{\the\scriptscriptfont0 #2}}

\def\gtorder{\mathrel{\raise.3ex\hbox{$>$}\mkern-14mu
             \lower0.6ex\hbox{$\sim$}}}
\def\ltorder{\mathrel{\raise.3ex\hbox{$<$}\mkern-14mu
             \lower0.6ex\hbox{$\sim$}}}

\def\semidirprod{\rlap{\ss C}\raise1pt\hbox{$\mkern.75mu\times$}}
\def\for{\lower6pt\hbox{$\Big|$}}
\def\fish{\kern-.25em{\phantom{abcde}\over \phantom{abcde}}\kern-.25em}


\def\boxit#1{\vbox{\hrule\hbox{\vrule\kern3pt
        \vbox{\kern3pt#1\kern3pt}\kern3pt\vrule}\hrule}}
\def\dalemb#1#2{{\vbox{\hrule height .#2pt
        \hbox{\vrule width.#2pt height#1pt \kern#1pt \vrule
                width.#2pt} \hrule height.#2pt}}}

\def\ol{\overline}
\def\frac#1#2{{{#1}\over{#2}}}

\def\noin{\noindent}


\def\cosech{{\rm cosech\,}}

\def\etc{{\it etc.}}
\def\viz{{\it viz.}}
\def\eg{{\it e.g.}}
\def\ie{{\it i.e. }}
\def\cf{{\it cf }}
\def\pa{\partial}

\def\ket#1{\mid#1\rangle}

\def\me#1#2#3{\langle{#1}\mid\!{#2}\!\mid{#3}\rangle}  


\def\sumdasht#1#2{{\mathop{{\sum}'}_{#1}^{#2}}}

\def\3j#1#2#3#4#5#6{\left\lgroup\matrix{#1&#2&#3\cr#4&#5&#6\cr}
\right\rgroup}

\def\man{{\cal M}}

\def\m?{\mgn{?}}

\def\pa{\partial}

\def\beq{\begin{eqnarray}}
\def\eeq{\end{eqnarray}}


\def\aop#1#2#3{{\it Ann. Phys.} {\bf {#1}} ({#2}) #3}

\def\cmp#1#2#3{{\it Comm. Math. Phys.} {\bf {#1}} ({#2}) #3}
\def\cqg#1#2#3{{\it Class.Quant.Grav.} {\bf {#1}} ({#2}) #3}

\def\jmp#1#2#3{{\it J. Math. Phys.} {\bf {#1}} ({#2}) #3}
\def\jpa#1#2#3{{\it J. Phys.} {\bf A{#1}} ({#2}) #3}

\def\np#1#2#3{{\it Nucl. Phys.} {\bf B{#1}} ({#2}) #3}

\def\prB#1#2#3{{\it Phys. Rev.} {\bf B{#1}} ({#2}) #3}
\def\prD#1#2#3{{\it Phys. Rev.} {\bf D{#1}} ({#2}) #3}
\def\prl#1#2#3{{\it Phys. Rev. Lett.} {\bf #1} ({#2}) #3}

\def\am#1#2#3{{\it Acta Mathematica} {\bf {#1}} ({#2}) #3}

\def\jpamt#1#2#3{{\it J. Phys.A:Math.Theor.} {\bf{#1}} ({#2}) #3}
\def\jram#1#2#3{{\it J. f. reine u. Angew. Math.} {\bf {#1}} ({#2}) #3}

\def\mz#1#2#3{{\it Math. Zeit.} {\bf {#1}} ({#2}) #3}

\def\plb#1#2#3{{\it Phys. Letts.} {\bf {B#1}} ({#2}) #3}

\def\qjm#1#2#3{{\it Quart. J. Math.} {\bf {#1}} ({#2}) #3}

\begin{title}
\vglue 0.5truein
\vskip15truept
\centertext {\Bigfonts \bf Quantum revivals in free field CFT} \vskip7truept
\vskip10truept\centertext{} \vskip17truept \centertext{\Bigfonts \bf }
 \vskip 20truept
\centertext{J.S.Dowker\footnote{ dowker@man.ac.uk;  dowkeruk@yahoo.co.uk}} \vskip
7truept \centertext{\it Theory Group,} \centertext{\it School of Physics and Astronomy,}
\centertext{\it The University of Manchester,} \centertext{\it Manchester, England} \vskip
7truept \centertext{}

\vskip 7truept  \vskip40truept
\begin{narrow}
A commentary is made on the recent work by Cardy, ArXiv:1603. 08267, on quantum
revivals and higher dimensional CFT. The actual expressions used here are those derived
some time ago. The calculation is extended to fermion fields for which the power spectrum
involves the odd divisor function. Comments are made on the equivalence of operator
counting and eigenvalue methods, which is quickly verified. An explanation of the rational
revivals for odd spheres is given in terms of wrongly quantised fields.

\end{narrow}
\vskip 5truept
\vskip 60truept
\vfil
\end{title}
\pagenum=0
\newpage

\section{\bf 1. Introduction and summary.}

In a discussion of the topic of quantum revivals, on quenching, in the context of conformal
field theory, Cardy, [\pref{Cardy2}], when extending his previous 2d analysis,
[\pref{Cardy1}], to higher dimensions, encounters, for free fields, what is, essentially,
finite temperature theory on a generalised Einstein universe, \ie a `generalised torus',
more particularly a generalised cylinder $I\times \cal M$. $I$ is an interval and $\cal M$
here is a $d$--sphere.\footnote{ My $d$ differs from that in [\pref{Cardy2}] by 1.}

The precise object sought is the return amplitude
$A(t,*)=\big|\me{\psi_0}{e^{-iHt}}{\psi_0}\big|$ for some quenching initial state,
$\ket{\psi_0}$. $A$ is a function of the evolution time, $t$, and whatever parameters
$\ket{\psi_0}$ depends on. $H$ is the field Hamiltonian.

For the particular, conformal choice of $\ket{\psi_0}=e^{-\be H/4}\ket D$, $\ket D$ being
a Dirichlet boundary state, $A(t,\be)$ is determined by the partition function, $Z$, on a
generalised cylinder, with Dirichlet conditions on the interval, $I(\be/2)=[0,\be/2]$. The
specific expression is,
  $$
  \log A(t,\be)={1\over2}\Real\bigg(
  \log Z\big({\cal M}\times I(\be/2+it)\big)-
  \log Z({\cal M}\times I(\be/2)\big)\bigg)\,.
  \eql{ret}
  $$
For fixed $\be$, the second term can be omitted for graphical purposes. The parameter
$\be$ is ultimately interpreted as an inverse temperature and I will formally treat it as
such.

I wish to draw attention to earlier calculations, [\pref{Apps,DandA1,DandA2}], of $\log Z$
as the log determinant of the relevant propagating operator, $\caO$, on the cylinder,
 $$
{1\over2}\log\det \caO=-{1\over 2}\ze'_\caO(0)\,.
\eql{logz}
 $$
I have used the \zf\ definition of a functional determinant.

The free energy has also been derived in [\pref{ChandD}].

In this paper, I present some technical remarks, in the free field setting, taking earlier work
into account, and make some connections which might be interesting.

The fundamental equations are recalled in section 2, closely following [\pref{Cardy2}], and
the return amplitude evaluated using earlier found expressions for the partition function.
The results, of course, are the same as in [\pref{Cardy2}]. Section 3 discusses the
consequences of modular invariance, introduced a little differently to [\pref{Cardy2}] with
the same conclusion but for all dimensions. In section 4,  these calculations are repeated
for the spin--half, fermion field, a case not considered in [\pref{Cardy2}]. The power
spectrum is given in section 5 and found to involve the odd divisor function. Section 6
treats the situation where the spatial  $d$--sphere is quotiented by a cyclic group. An
explanation of the sign reversed revivals using `wrongly quantised' fields is made in
section 7. In Appendix A, the equivalence of the operator counting and eigenvalue
methods is verified and Appendix B derives expressions for boson and fermion
entanglement entropies.

\section{\bf2. The calculations}
The first step in the evaluation of the determinant was to rearrange the interval modes. In
[\pref{Apps,DandA1,DandA2}] this was neatly expressed in \zf\ terms as
  $$
  \ze({I\times\cal M})={1\over2}\big(\ze(S^1\times\man)\mp \ze(\man)\big)\,,
  $$
where the $\pm$ gives Neumann and Dirichlet conditions on the interval. Inserted into
(\peq{ret}) the second term cancels, being $\be$ independent. This conclusion is also
reached by Cardy.

Equation (\peq{ret}) can therefore be replaced by
  $$
  \log A(t,\be)={1\over4}\Real\bigg(
  \log Z\big({\cal M}\times S^1(\be/2+it)\big)-
  \log Z({\cal M}\times S^1(\be/2)\big)\bigg)\,.
  \eql{ret2}
  $$i

As noted and used in [\pref{Apps,DandA1,DandA2}], the main problem then reduces to a
thermal one on the Einstein universe, which is a topic with a history, (\eg\
[\pref{{DandC},{Kennedy},{Unwin1}, {AandD}}]). $S^1$ can be referred to as the
thermal circle. Consult also [\pref{CapandC}].

In general, in \zf\ regularisation on any manifold, the effective action, essentially $\log
Z$, consists of a divergence, with an associated logarithm, plus a finite part which has the
form (\peq{logz}).

The divergence and logarithmic parts are controlled by the conformal anomaly on $\cal
M\times I$ and a simple argument shows that for conformal coupling (for odd dimensional
spheres) this anomaly is zero. It is automatically zero for even spheres.

This being so, on the basis of \zf\ regularisation on can set, in (\peq{ret2}), for bosons,
  $$
  \log Z={1\over 2}\ze'_\caO(0)\,,
  $$
where $\caO$ is the thermal propagation operator.

Notationally, from now on, in order to avoid confusion, I set $\Xi=\log Z$. This includes the
zero temperature part, $\Xi_0$, and totally $\Xi=\Xi_0+\Xi'$, $\Xi'$ being the finite
temperature correction. The reason for this is that in [\pref{Cardy2}] and elsewhere, $Z$
refers to the usual partition function, the sum over Fock space states \ie $\log Z$ is just
$\Xi'$. Actually it does not matter whether $\Xi$ or $\Xi'$ is used in (\peq{ret}) as $\Xi_0$
cancels on taking the difference and real part.\footnote{ This being the case, one does not
really need the full apparatus of \zf\ regularisation.}

The general  statistical sum eqn.(31) or, equivalently (55), in [\pref{DandK}] can be
written to give
  $$
  \be F= \be E_0-\sum_{m=1}^\infty {1\over m} K^{1/2}(m\be)\,.
  \eql{eff}
  $$
$F$ is the conventional free energy, $\be F=-\Xi$ and $K^{1/2}$ is the degeneracy
generating function or, equivalently, the `cylinder kernel' for the pseudo operator
(Hamiltonian) $\sqrt D$, $D$ being the propagating operator on $\man$. In the present
instance this is related to the conformally invariant (Yamabe--Penrose) Laplacian, $Y$, by
$D=Y+1/4$. $K^{1/2}$ is a single particle sum--over--states partition function. $E_0$ is
the zero temperature, Casimir energy.

For the present spherical situation, the expression is given explicitly in [\pref{ChandD}],
equn.(78), for $\man$ an orbifold quotient of the sphere, S$^d$, in particular for a
hemisphere, and thence, by addition, for a full sphere. The expressions for this latter case
can also, conveniently, just be read off from [\pref{Apps,DandA1,DandA2}].

Since, notationally, I  am generally adhering to [\pref{Cardy2}], I give the relation with
the parameters used in [\pref{Apps,DandA1,DandA2}] (shown first),
  $$
  L=\be/2\,,\quad a=L/2\pi\,.
  $$
$a$ is the sphere radius. In order to simplify the exposition, I set now $a=1$ \ie
$L=2\pi$.

The expressions given in [\pref{Apps,DandA1,DandA2}] and [\pref{ChandD}] imply that,
\footnote{ An outline of the derivation is given in Appendix A.}
  $$
  \Xi_d'(\be)={1\over 2^d}\sum_{m=1}^\infty{1\over m}\cosh (m\be/2)
  \,\cosech^d (m\be/2)\,,
  \eql{ssum}
  $$
which holds for odd and even sphere dimensions.

As remarked, it is sufficient to use (\peq{ssum}) to display the return amplitude,
(\peq{ret}). Figs. 1 and 2 plot $\Xi'_d(\be+2i\pi s)$ for $d=2$ and $d=3$. $s$ equals
$t/\pi$.

\epsfxsize=5truein \epsfbox{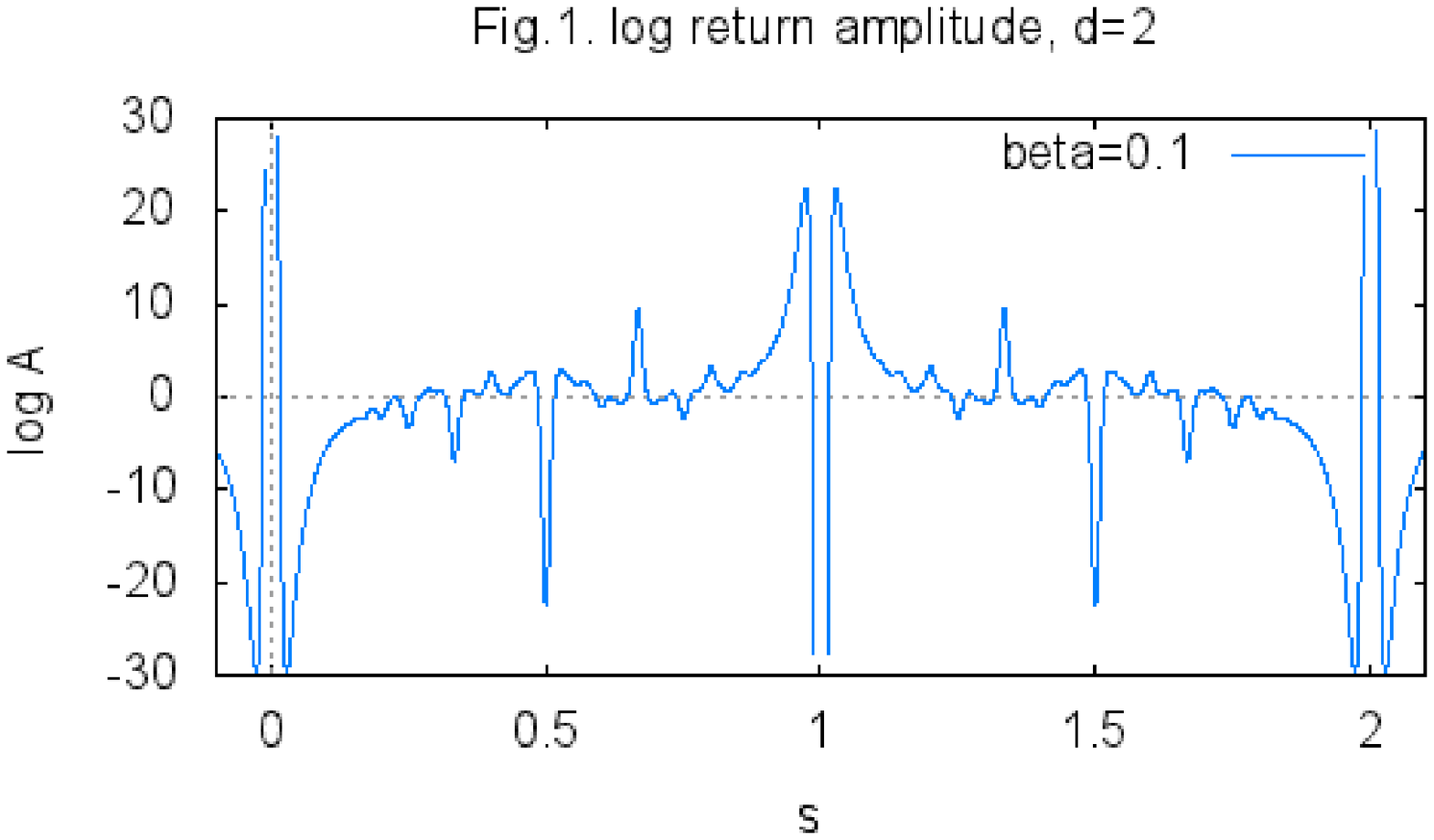}

\epsfxsize=5truein \epsfbox{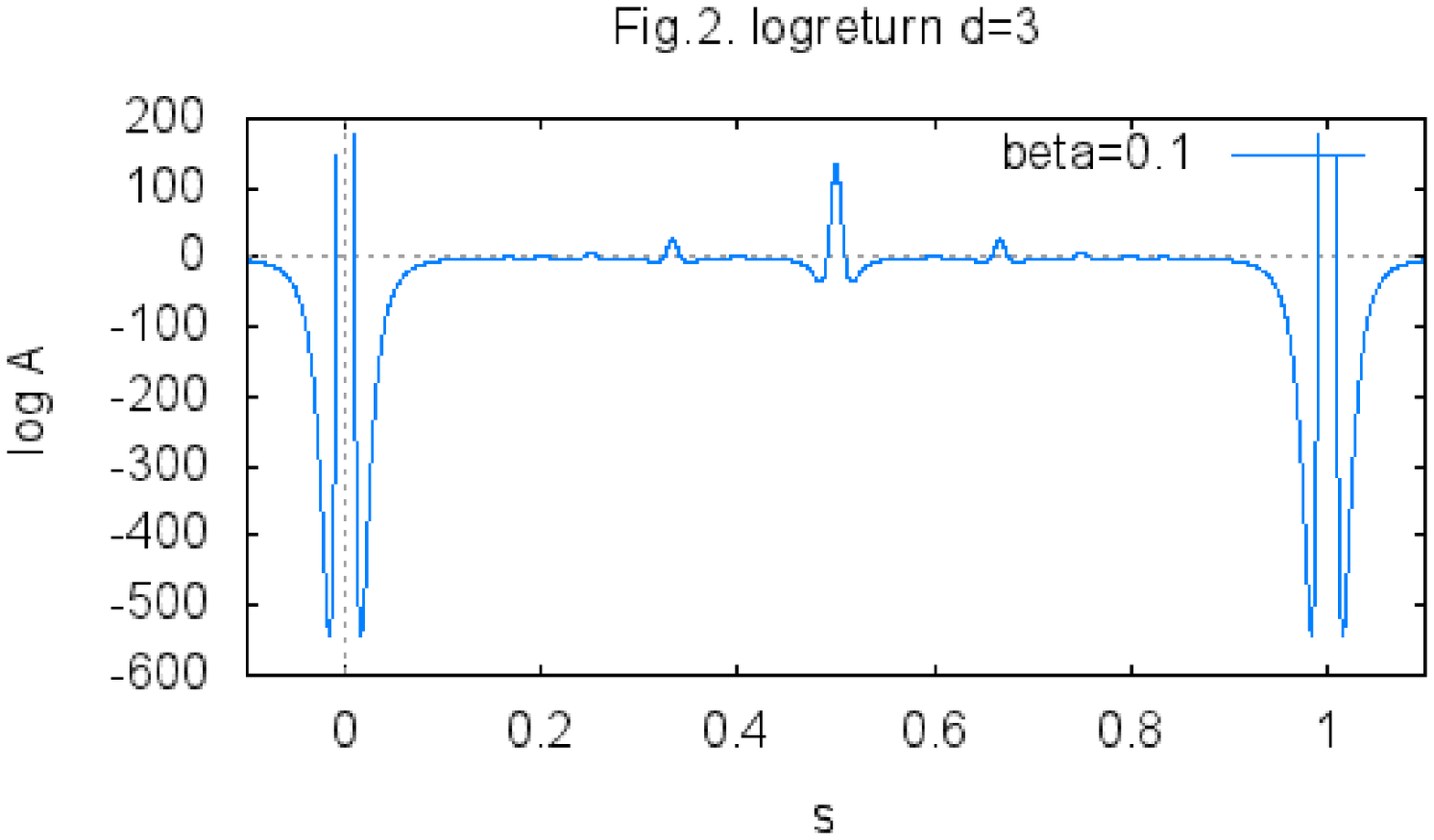}

These are both plotted in [\pref{Cardy2}] but I have extended the horizontal range a little
in order to accentuate any periodicity.

\begin{ignore}
\section{\bf Twisted fields}

Fields on the thermal circle can be twisted, for example real fields can change sign on
going once around the circle. In this case I will go through some of the formalism
telescoped in the previous section.

I define a twisted thermal \zf\  on ${\tilde S}^1\times S^d$ by
  $$
  {\tilde\ze}_\caO(s)=\sum_{n=-\infty}^{\infty}\sum_{l=0}^\infty
  {g_l\over\big((l+(d-1)/2)^2/a^2+4\pi^2(n+1/2)^2/b^2\big)^s}
  $$
where $g_l$ is the degeneracy of the scalar conformal eigenlevel on S$^d$, labelled by
$l$. I do not need its form.

As usual, the summation over odd integers is re--expressed. Introducing the untwisted \zf,
  $$
   \ze_\caO(s,b)=\sum_{n=-\infty}^{\infty}\sum_{l=0}^\infty
  {g_l\over\big((l+(d-1)/2)^2/a^2+4a^2\pi^2 n^2/b^2\big)^s}
  $$
one has a familiar relation,
  $$
  {\tilde\ze}_\caO(s)=a^{2s}\big[\ze_\caO(s,2b)-\ze_\caO(s,b)]
  $$
and the previous results can be combined to give
  $$\eqalign{
  {\log {\tilde Z}}_2&=S_2\big({2\pi\be\over L}\big)-S_2\big({\pi\be\over L}\big)\cr
  {\log Z}_3&=S_3\big({\pi\be\over L}\big)-{\pi\be\over{120 L}}\cr
  {\log Z}_4&=S_4\big({\pi\be\over L}\big)\cr
  {\log Z}_5&=S_5\big({\pi\be\over L}\big)+{31\pi\be\over{30240 L}}\cr
  }
  $$
\end{ignore}

\section{\bf 3. Modular invariance}

Cardy relates the partial revivals at rational values of $s$, evidenced by the maxima in the
curves, to the modular properties of the free energy. Actually, it is the internal energy
(including the Casimir term) that has the simpler behaviour (for odd $d$). In the Einstein
universe, this was early recognised, [\pref{DandC}], [\pref{CandD}], and also was
discussed by Cardy [\pref{Cardy3}] in a conformal field theory context, specific higher
dimensions being considered. A few comments occur in [\pref{Dowzerom}] and a more
extensive analysis was given in [\pref{DandKi}] for odd dimensional spheres. Further
calculations are provided in [\pref{GPP}] for different field contents and an extension to
$AdS_d$.

I recapitulate a few details of [\pref{DandKi}]. The mode structure (eigenlevels and their
degeneracies) on spheres is an ancient topic and needs no explanation. The upshot is
that, in the usual description, the conformal eigenlevels are squares of integers, say
$n^2$, with degeneracies  that are polynomials in $n^2$.

It is best, for present purposes, to treat the terms in this polynomial individually and
then, if required, reconstitute the full expression. Selecting the $(2l-2)$ power, the
standard statistical formula gives for the corresponding `partial' energy
  $$\eqalign{
  \ep_l(\xi)&=\ep_l(0)+\sum_{n=1}^\infty {n^{2l-1}q^{2n}\over1-q^{2n}}\cr
  &\equiv \ep_l(0)+\ep'_l(\xi)\,.
  }
  \eql{paren}
  $$
$\ep'$ is the energy finite temperature correction.

I have rescaled the energy by the sphere radius, $a$, ($=1$ here) introduced the
dimensionless parameter $\xi=2\pi a/\be$ and defined $q=\exp(-\pi/\xi)$. The quantity
$\ep_l(0)$ is the (scaled) partial Casimir energy. Explicitly,
  $$
  \ep_l(0)=-{\caB_{2l}\over 4l}\,,
  \eql{vacen}
  $$
in terms of Bernoulli numbers, $\caB$.

The easiest way of showing the inversion symmetry,
  $$
  {1\over\xi^l}\,\ep_l(\xi)=(-1)^l \xi^l\,\ep_l(1/\xi)\,,
  \eql{inv}
  $$
is to relate $\ep$ to a (holomorphic) Eisenstein series,
  $$
G_l(\om_1,\om_2)\equiv
\sumdasht{m_1,m_2=-\infty}
\infty{1\over \big(m_1\om_1+m_2\om_2\big)^{2l}}\,,
\eql{eis}
  $$
by
  $$
  \ep_l(\xi)=(-1)^l\,C(l)\,G_l(1,i/\xi)
  $$
($C(l)$ is an inessential constant). This connection is a basic result in analytic number
theory. The inversion symmetry, (\peq{inv}), follows immediately. More generally, the
expression is invariant under the modular group action on the periods $\om_1,\,\om_2$.
The translational generator is $\sssb\to \sssb-i$ with, for convenience and to agree with a
previous notation, [\pref{Dowmod}], I have put $\sssb\equiv\xi^{-1}$.

Note that the Casimir energy appears naturally. If it is extracted according to
(\peq{paren}), equation (\peq{inv}) reads
  $$
  {1\over\xi^l}\big(\ep_l(0)+\ep'_l(\xi)\big)=
  (-1)^l\xi^l\big(\ep_l(0)+\ep_l'(1/\xi)\big)\,.
  \eql{sinv}
  $$
In this form, the identity can be traced back at least to Ramanujan, see [\pref{DandKi}].

To relate the high and low temperature regimes, let $\xi$ become large in (\peq{sinv}).
From its form, $\ep'(1/\xi)$ tends to zero exponentially fast\footnote{ This  is generally
true, for finite systems, [\pref{DandK}].} and so, up to the such terms,
  $$
  \ep_l'(\xi)+\ep_l(0)\approx(-1)^l\,\ep_l(0)\xi^{2l}\equiv\si_l\,\xi^{2l}\,\sim \si_l T^{2l}
  \eql{hitemp}
  $$
connecting high temperature on the left, to low temperature (the $\ep_l(0)$) on the right.
The right--hand side is the (partial) Planck term \footnote{ This corresponds to the Weyl
universal term in the asymptotic distribution of eigenvalues.} and $\si_l$ is a (partial)
Stefan--Boltzmann constant, which is positive, using (\peq{vacen}).

For the $d$--dimensional sphere, the actual energy, $aE$, is a sum of $\ep_l$ where $l$
runs from 2 to $(d+1)/2$ and the high temperature behaviour is a sum of terms like
(\peq{hitemp}). That there are only a finite number agrees with the general expression for
the high temperature limit in terms of the heat--kernel coefficients, [\pref{DandK}],
[\pref{Dowfint}]. This is because the conformal heat--kernel expansion terminates on odd
spheres (up to exponential corrections). The inversion properties of $E$ are therefore not
straightforwardly expressed, apart from the three--sphere when $l=2$. However for many
purposes, the dominant term is given by $l=(d+1)/2$ and suffices.

More logically, the expansion for the free energy ($\be F=-\Xi$) would be derived, as in
[\pref{DandK}], from first principles, the energy following by differentiation ($E=\pa (\be
F)/\pa \be$). To utilise  the inversion behaviour, (\peq{inv}), this procedure is reversed
(\cf\ [\pref{Cardy2}]).

According to the expression (\peq{ret}) for the return amplitude, one needs to make the
replacement $\xi^{-1}\to\xi^{-1}+is$ in the thermodynamic quantities. The analysis is
eased by taking $\xi$ large. Then the revivals at rational $s$ are more pronounced
because the initial value (the Planck term) is large when $\xi\to\infty$ and $s\sim0$. In
this case it is more convenient to use $\sssb$. The standard Planck contribution to the
free energy then says that the initial (partial) log return amplitude, (\peq{ret}), is
  $$
  \log A_l(s)\approx  C_l\,\Real {1\over (\sssb+is)^{2l-1}}\,,\quad s \,\,\,{\rm small}\,.
  \eql{initial}
  $$
where the constant $C_l=2\pi\si_l/(2l-1)$.

The total amplitude is a linear combination of the $\log A_l$. For small enough $\sssb$
and $s$ this is dominated by the first term, \ie the one with the largest $l,\,=(d+1)/2$.
\footnote{ It is conventional, in a general manifold, to refer to just this term as {\it the}
Planck term although it is not strictly thermal. Cardy refers to it as the Casimir term} It
can be checked graphically that this gives a good approximation to the complete quantity
obtained from (\peq{ssum}).

Because of the translational invariance under $\sssb\to\sssb\pm i$, there will be identical
copies of the initial behaviour, (\peq{initial}),  around integral $s$. Multiple translations,
combined with inversion replicates this behaviour, with reduced amplitude, at rational $s$.
The argument is as follows, [\pref{Cardy2}].

The modular relation, (\peq{sinv}), is written in terms of the partial finite temperature
correction part of $\Xi$, $\Xi'_l$, which is the quantity plotted. Then,
  $$
  {\pa\over \pa\xi}\,\Xi_l'(\xi)=(-1)^{l+1} \xi^{2l-2}\,{\pa\over \pa\xi}\,\Xi_l'(1/\xi)
  +(-1)^l 2\pi\ep_l(0)\,\big((-1)^l\xi^l-\xi^{-l}\big)\,.
  \eql{mod2}
  $$

The simplest case is one inversion combined with a translation of $\xi^{-1}$  by $im$,
with $m$ integral. This converts the region around $s=1/m$ to that around $s=0$. The
first region is accessed by  setting $\xi^{-1}\approx{\be\over2\pi}+i/m+i\ep$ with $\ep$
(and $\be$) small. One requires the left--hand side in this region. This is provided by the
right--hand side. Then one has $\xi\approx -im+{\be\over2\pi}m^2+i\ep\,m^2$.
Translating away the $-im$ gives $\xi\to m^2(\be/2\pi+i\ep)$. Since this is small, the
right--hand side is well approximated by the high temperature form (\peq{hitemp}), on
ignoring the last term. Hence
  $$
  \Xi_l'\bigg|_{s=1/m+\ep}\sim(-1)^{l-1}
  (-im)^{2l-2}{1\over(\sssb+i\ep)^{2l-1}m^{4l-2}}\,.
  $$

This shows that the profile around $s=1/m$ is the same as that around $s=0$, except for
a reduction in amplitude by a factor of $1/m^{2l}$. For the dominant term this equals
$1/m^{d+1}$ which can be checked numerically from the complete expression,
(\peq{ssum}).

Repeating this procedure, [\pref{Cardy1}], allows the revivals at the rational points $n/m$
to be obtained The result is the same, with the amplitude reduction still $1/m^{d+1}$.

This analysis is just for odd spheres, but the complete expressions, (\peq{ssum}), hold
for {\it all} $d$ and it is these that have been plotted. It is noticed empirically that there
are also revivals for even $d$, but the structure is not straighforwardly analysed.

\section{\bf4. Spin--half}

The analysis can be repeated for spin-1/2 fields. Mixed boundary conditions are
conformally  invariant.  There are two such boundary conditions, which yield identical
spectral results and are effectively fermionic (\ie antisymmetric on the thermal circle). The
relevant formulae on the generalised cylinder and torus are given in [\pref{Apps}],
[\pref{DandA1}].

The fermion effective action leads to,
    $$
    \Xi^f_d=\be\,E_0^f+{\Xi^f_d}'\big({\be}\big)\,,
    \eql{logzf}
    $$
where $E_0^f$ is the classic Dirac Casimir energy on the $d$--sphere and the correction
${\Xi^f_d}'$ is found to be,
  $$\eqalign{
  {\Xi_d^f}'(\be)&=-{2\over2^{[(d-1)/2]}}\sum_{m=1}^\infty(-1)^m
  {1\over m}\cosech^d(m\be/2)\cr
  &=
  -{2\over2^{[(d-1)/2]}}\sum_{m=1}^\infty{1\over m}\bigg(\cosech^d(m\be)-
  \cosech^d(m\be/2)\bigg)\,.
  }
  \eql{ssf}
  $$

As explained before, just the second term on the right--hand side of (\peq{logzf}) is
plotted. As examples, revivals, partial and complete, can be seen for the $d=1$, $d=2$
and $d=3$ cases shown in figs.3, 4 and 5.

\epsfxsize=5truein \epsfbox{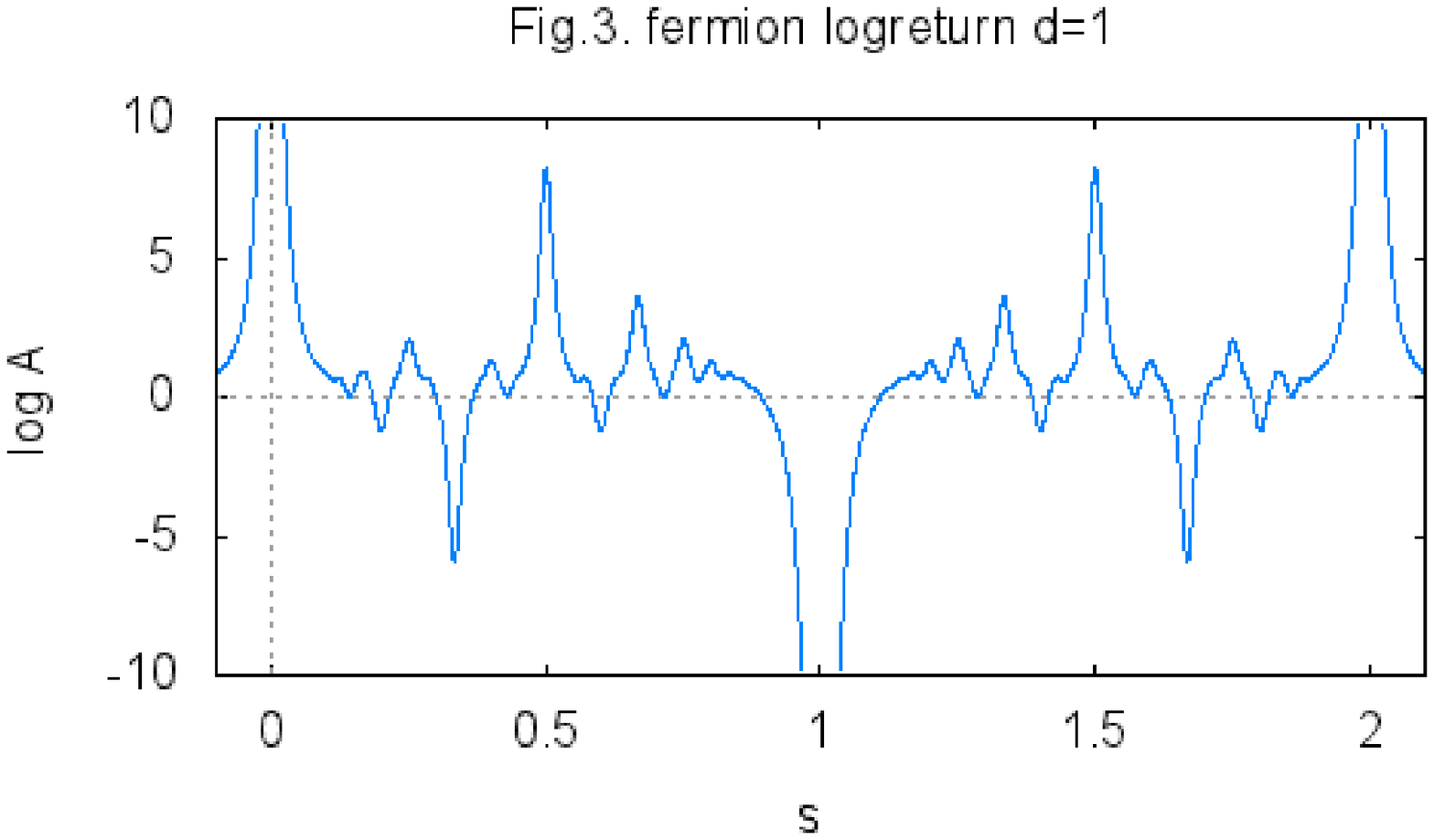}

\epsfxsize=5truein \epsfbox{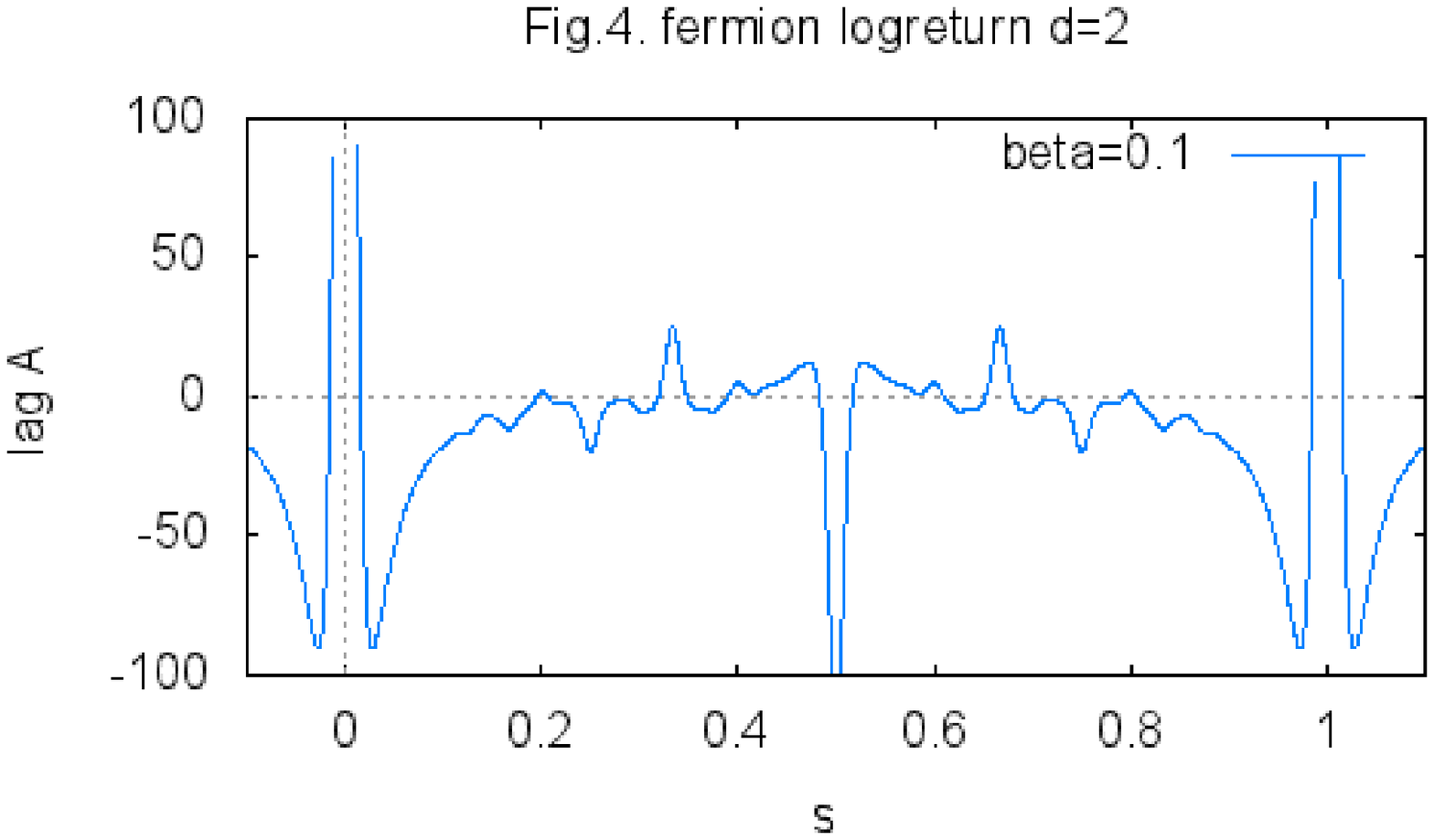}

\epsfxsize=5truein \epsfbox{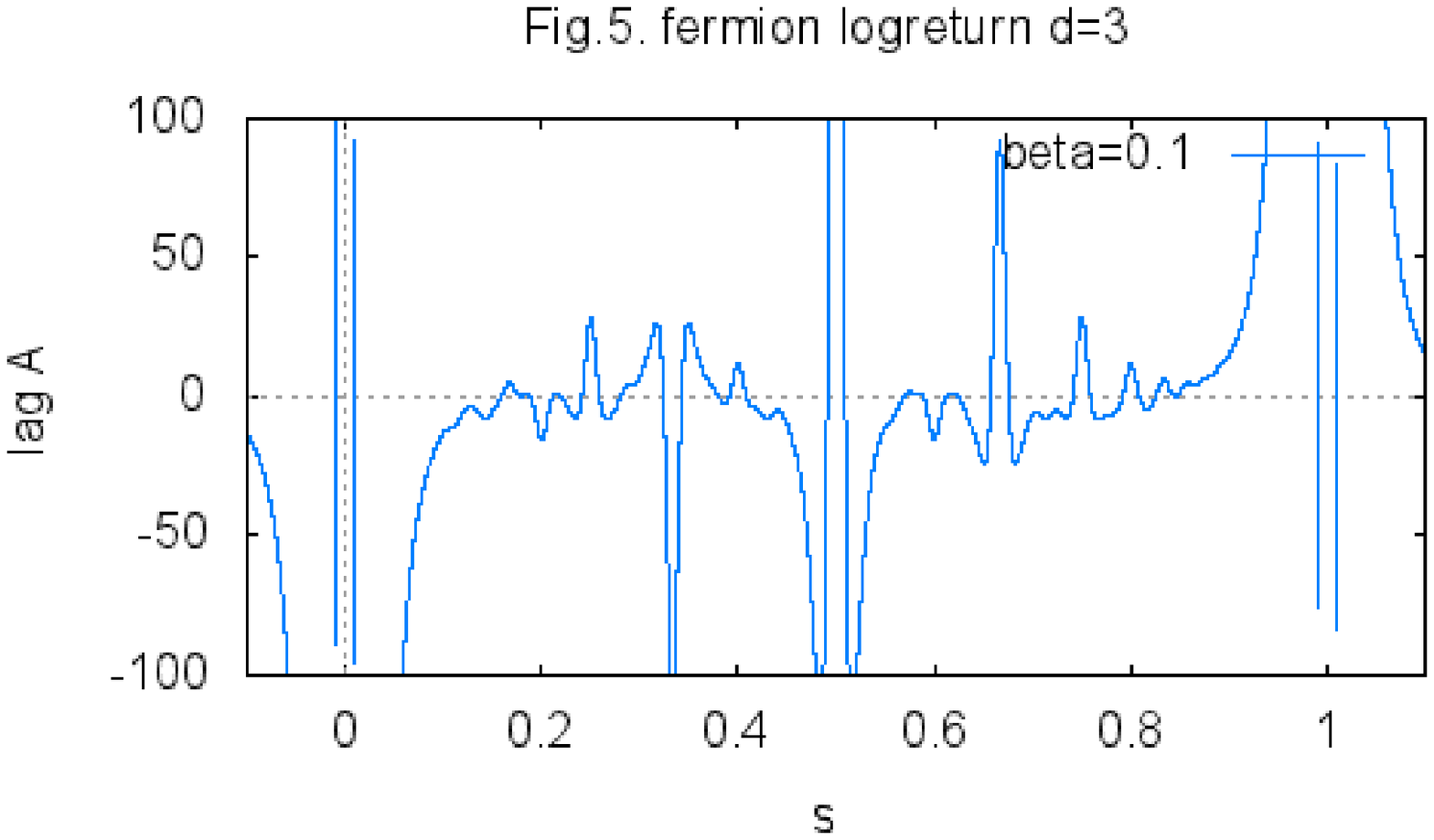}

As for the scalar field, these can be traced to the modular invariance of the spinor partial
total internal energy, $\eta_l(\xi)$, which is related, for odd $d$, to a doubly
sign--modulated Eisenstein series,
  $$
H_l(\om_1,\om_2)\equiv
\sumdasht{m_1,m_2=-\infty}
\infty{(-1)^{m_1+m_2}\over \big(m_1\om_1+m_2\om_2\big)^{2l}}\,,
\eql{eisf}
  $$
through, [\pref{DandKi}],
  $$
  \eta_l(\xi)=(-1)^{l+1}\,C_f(l)\,H_l(1,i/\xi)\,.
  $$
$C_f$ is an unrequired constant.

$\eta_l(\xi)$ therefore enjoys the same inversion properties, for odd spheres, as the scalar
quantity, $\ep_l(\xi)$. The translation behaviour is altered because of the twisting. For
odd spheres the periodicity is now $2$. This also follows from the total expression
(\peq{ssf}). In fig.5, the complete period is obtained by reflecting in the line, $s=1$.

Because of this periodicity, the attenuations  to $s=1/m$ from $s=0$ , for $m$ even, and
from $s=1$, for $m$ odd, both equal $1/m^{d+1}$. A more extended analysis is
presented in section 7.

The maximum at $s=1$ can be investigated from the forms (\peq{ssf}) which show, in
addition, that the periodicity is 1 for even spheres.

The underlying mechanism giving rise to revivals at the rationals for even spheres has yet
to be elucidated.
\section{\bf5. The fermion power spectrum}

Since Cardy provides a treatment of the scalar case, I need present only the fermion
analysis.

Working in terms of the partial quantities (therefore only odd spheres are covered) the
Fourier series form of the $q$--series for $\eta_l$ is standard elliptic fare. Glaisher,
[\pref{Glaisher1,Glaisher2}], conveniently has the requisite lists. The fermion series is,
(see [\pref{DandKi}] equn.(20)),
  $$
  \sum_{n=0}^\infty (2n+1)^{2l-1} {q^{2n+1}\over1+q^{2n+1}}
  =\sum_{n=1}^\infty(-1)^{n-1}\De_{2l-1}(n)\,q^n\,,
  \eql{ft}
  $$
where $\De_k(n)$ is the odd divisor function related to the usual one, $\si_k$, by
   $$
   \De_k(n)=\si_k(n)-2^k\si_k(n/2)\,,
   $$
with $\si_k$ at a half--integer defined zero.

In the case under consideration, $q$ takes the form $q=e^{-\be/2+i\pi s}$.

Equation (\peq{ft}) refers to the energy. To find ${\Xi^f}'$ an integration with respect to
$\be$ yields the factor $2/n$. Then, taking the real part gives,
  $$
  \Real {\Xi^f}'_l\approx2\sum_{n=1}^\infty(-1)^{n-1}{1\over n}\De_{2l-1}(n)\,
  e^{-\be n/2}\cos(\pi n s)\,,
  $$
implying the power spectrum amplitude,
  $$
  {2\over n}\De_{2l-1}(n)\,e^{-\be n/2}\,.
  $$

\epsfxsize=5truein \epsfbox{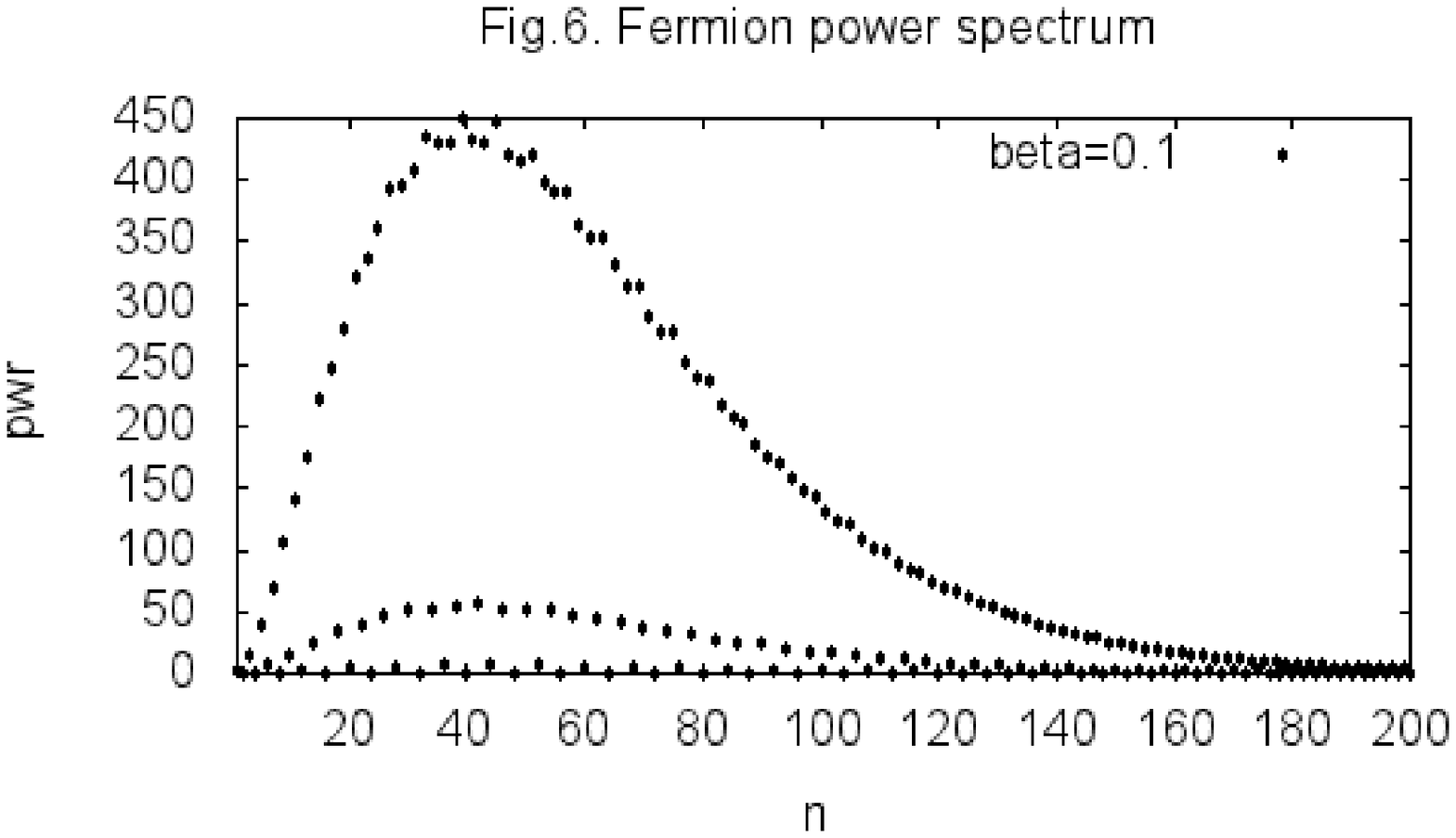}

Fig.6 plots this out for $l=2$, corresponding to the three--sphere. The apparently different
curves are due to the behaviour of the odd divisor function.

$n$ measures the frequency in units of $1$. Even frequencies are suppressed, some
severely. The curves for the higher spheres are similar in shape but more extreme.
\section{\bf6.  Spherical factors}

Taking quotients of the sphere does not destroy any conformal properties and one can
pursue the same path to find the return amplitude. However, the exact inversion behaviour
is lost, [\pref{DandKi2}].

The free energy, and hence $\Xi$, were given in [\pref{ChandD}] for the quotients by a
regular solid symmetry group, $\Ga$. In particular for the cyclic case
($\Ga=\oZ_B,\,B\in\oZ$) the formula easily gives, see also [\pref{DandKi2}],
  $$
  \Xi_d(\be,B)=-\be E_{d,0} (B)+\Xi'_d(\be,B)\,,
  $$
where,\mgn{Check 2s}
  $$
  \Xi'_d(\be,B)={1\over2^{d}}\sum_{m=1}^\infty{1\over m}
  \coth(mB\be/2)\,\cosech^{d-1}(m\be/2)\,.
  $$

\epsfxsize=5truein \epsfbox{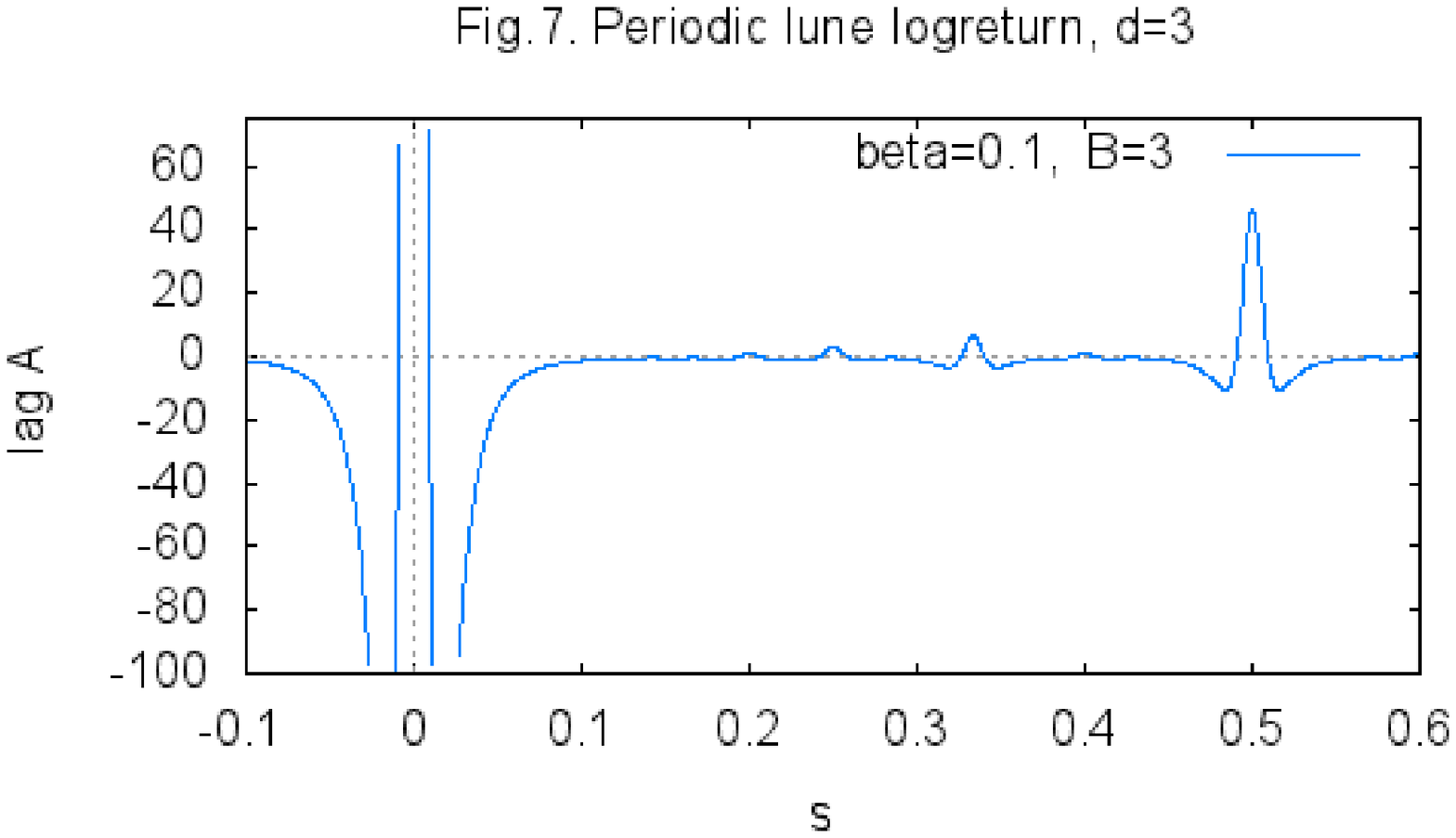}

Again, the Casimir term, $E_0$, can be ignored for plotting and, as an example, the log
return for the odd dimensional $B=3$ periodic lune is shown in fig.7. The period is 1 and
there are returns at the rationals, just as for the full sphere, $B=1$. Only the  vertical
scale changes as $B$ varies, despite the lack of inversion symmetry for $B\ne1$. In fact
the attenuation factor is independent of $B$.

Taking the $B\to\infty$ limit one finds
$$\eqalign{
  \Xi'_d(\be,\infty)&={1\over2^{d}}\sum_{m=1}^\infty{1\over m}
  \cosech^{d-1}(m\be/2)\,,\cr
  }
  $$
which is, to a factor, the $\Xi'$ for a bosonic spinor in one dimension less. (See next
section).

\section{\bf 7. Spin--zero fermions and spin--half bosons}
Cardy, [\pref{Cardy2}], explained the integer revivals at $s=1,3,\ldots$ for even spheres,
Fig.1, in terms of a fermionic scalar $\log Z$. A more extensive treatment is possible for
odd spheres as I now show.

First I present some facts which result from an inspection of Glaisher's tables in
[\pref{Glaisher2}]. Referring to that on p.64, I compare the $q$--series second from top
with the one second from bottom, \viz,
  $$\eqalign{
  X'(q)=&\sum_1^\infty {(2n)^{2l-1} q^{2n}\over1+q^{2n}}\cr
  Y'(q)=&\sum_0^\infty {(2n+1)^{2l-1} q^{2n+1}\over1-q^{2n+1}}\,.
  }
  $$

$X'$ corresponds to the (partial) internal energy correction of a kinematic spin--zero field
thermalised as a fermion, and $Y'$ to that for a kinematic spin--half field thermalised as a
boson.

For convenience, I give the corresponding standard boson and fermion series discussed
above (as in (\peq{paren}) and (\peq{ft})),
  $$\eqalign{
  B'(q)=&\sum_1^\infty {(2n)^{2l-1} q^{2n}\over1-q^{2n}}\cr
  F'(q)=&\sum_0^\infty {(2n+1)^{2l-1} q^{2n+1}\over1+q^{2n+1}}\,.
  }
  $$
Under translation, $q\to-q$, it is easily seen that $F'(-q)=-Y'(q)$. This explains the
inverted revivals depicted in fig.5 as I now show in detail.

The modular behaviour under inversion ($q\to \tilde q$ with $\log q\log\tilde q={\pi^2}$)
can be determined from the double sum representations given in the table. These are
singly twisted Eisenstein series for $X'$ and $Y'$.

Glaisher's manipulations allow one to write, ($q=e^{-\mu}$),
  $$\eqalign{
-{(2l-1)!\over4}\sum_{-\infty}^\infty\sum_{-\infty}^\infty {(-1)^s\over
(s\mu+ri\pi)^{2l}}&=X'_l(q)+2^{2l-1}{\caB_{2l}\over4 l}
={\ep'^{(fs)}(\xi)+\ep^{(fs)}(0)\over2^{1-2l}}\cr
{(2l-1)!\over4}\sum_{-\infty}^\infty\sum_{-\infty}^\infty {(-1)^r\over
(s\mu+ri\pi)^{2l}}&=Y'_l(q)+(2^{2l-1}-1){\caB_{2l}\over4 l}
={\ep'^{(bd)}(\xi)}+{\ep^{(bd)}(0)}\cr
-{(2l-1)!\over4}\sum_{-\infty}^\infty\sum_{-\infty}^\infty {(-1)^{s+r}\over
(s\mu+ri\pi)^{2l}}&=F'_l(q)-(2^{2l-1}-1){\caB_{2l}\over4 l}
={\ep'^{(fd)}(\xi)}+{\ep^{(fd)}(0)}\cr
{(2l-1)!\over4}\sum_{-\infty}^\infty\sum_{-\infty}^\infty {1\over
(s\mu+ri\pi)^{2l}}&=B'_l(q)-2^{2l-1}{\caB_{2l}\over 4l}
={\ep'^{(bs)}(\xi)+\ep^{(bs)}(0)\over2^{1-2l}}\cr
}
\eql{fsbs}
  $$
To accord with my previous notation I have $\mu=\pi/\xi$ and on the right have dropped
the $l$ label to avoid overload. Also $f$ means fermionic, $b$ bosonic, $s$ scalar and $d$
Dirac spinor.

As the notation indicates, $2^{1-2l} B'$ is the finite temperature correction part, $\ep'$,
of the (partial) internal energy for the conventional scalar, see (\peq{paren}). $F'$ is the
corresponding quantity, $\eta'$, for the conventional spinor. Similarly $2^{1-2l}X'$ and
$Y'$ are the finite temperature corrections for an anti--commuting scalar and commuting
spinor, respectively. The quantities on the right--hand side of (\peq{fsbs}) are, to a
factor, the total internal energies and it is these that satisfy the simplest modular
behaviour, as I show below. I denote them by $X,Y,F$ and $B$. The vacuum energies of
normal and odd fields with the same kinematics are equal and opposite, as expected.

As said, the double (Eisenstein) series allow the modular properties of these partial
quantities under inversion $\xi\to\xi^{-1}$, \ie $\mu\to\tilde\mu$ ,
$\mu\tilde\mu=\pi^2$, to be determined. Set $\tilde q=e^{-\tilde \mu}$. The inversion
behaviours of $B_l$ and $F_l$ have already been found, [\pref{DandKi}]. Now consider
that for $X_l$,

  $$\eqalign{
X'_l(\tilde q)+2^{2l-1}{\caB_{2l}\over4 l}&=-{(2l-1)!\over4}
\sum_{-\infty}^\infty\sum_{-\infty}^\infty {(-1)^s\over(s\tilde\mu+ri\pi)^{2l}}\cr
&=-{(2l-1)!\over4}
\sum_{-\infty}^\infty\sum_{-\infty}^\infty {(-1)^s\over(s\pi^2/\mu+ri\pi)^{2l}}\cr
&=-{(2l-1)!\over4}\bigg({\mu\over\pi}\bigg)^{2l}
\sum_{-\infty}^\infty\sum_{-\infty}^\infty {(-1)^s\over(s\pi+ri\mu)^{2l}}\cr
&=-{(2l-1)!\over4}\bigg({\mu\over\pi}\bigg)^{2l}
(-1)^l\sum_{-\infty}^\infty\sum_{-\infty}^\infty {(-1)^s\over(si\pi+r\mu)^{2l}}\cr
&=-{(2l-1)!\over4}\bigg({\mu\over\pi}\bigg)^{2l}
(-1)^l\sum_{-\infty}^\infty\sum_{-\infty}^\infty {(-1)^r\over(ri\pi+s\mu)^{2l}}\cr
&=-(-1)^l\bigg({\mu\over\pi}\bigg)^{2l}\bigg[Y'_l(q)+(2^{2l-1}-1){\caB_{2l}\over4 l}\bigg]
}
\eql{xtoy}
  $$

Therefore the finite temperature corrections are related by,\footnote{ This is one of  class
of identities referred to by Berndt, [\pref{Berndt}], as `hybrid'. See his Theorem 5.6.}
  $$'
 \bigg({\pi\over\mu}\bigg)^lX'_l(\tilde q)=-\bigg(-{\mu\over\pi}\bigg)^lY_l(q)-\bigg[2^{2l-1}
\bigg({\pi\over \mu}\bigg)^l+\bigg({\mu\over\pi}\bigg)^l(2^{2l-1}-1)\bigg]
{\caB_{2l}\over4 l}
\eql{invers2}
  $$
or
$$
 \xi^lX'_l(\xi^{-1})=-(-1)^l{1\over\xi^l}Y'_l(\xi)-\bigg[2^{2l-1}
\xi^l+(-1)^l{1\over\xi^l}(2^{2l-1}-1)\bigg]{\caB_{2l}\over4 l}\,,
\eql{invers3}
  $$
where I have  reverted to $\xi=\pi/\mu$ and, for ease, use $X_l(\xi)\equiv X_l(q)$ \etc

The effect of translations ($q\to-q$ or $1/\xi\to1/\xi+i$ or $\mu\to \mu+\pi i$ or $s\to
s+1$) follows easily from the $q$--series
  $$
  B_l(-q)=B_l(q)\,,\quad X_l(-q)=X_l(q)\,\quad {\rm and}\quad F_l(-q)=-Y_l(q)\,.
  \eql{trans}
   $$
The most interesting is the final transformation which interchanges commuting and
anti--commuting spinors with a sign change.

As before, the inversion relation, (\peq{invers3}), leads to the high temperature behaviour
of $X$ and $Y$. Letting $\xi\to \infty$ and $\xi\to0$ in turn gives,
  $$\eqalign{
   Y'_l(\xi)\to -(-1)^l\,2^{2l-1}{\caB_{2l}\over4l}\xi^{2l}\cr
   X'_l(\xi)\to-(-1)^l\,(2^{2l-1}-1){\caB_{2l}\over4l}\xi^{2l}\,.
  }
  $$

These results explain the revivals at the rationals for the fermion plots. The mechanism
runs exactly parallel to that given by Cardy, [\pref{Cardy2}], which was outlined above,
and the details need not be repeated. Taking the revivals at $s=1/m$, as before, the
explanation involves an inversion, followed by a translation through $m$. It is easiest to
treat $m$ even and $m$ odd separately for then the relevant translation in ({\peq{trans})
reads,
  $$
  F_l\big((-1)^{2e}\,q)=F_l(q)\,,\quad F_l\big((-1)^{2e+1}\,q)=-Y_l(q)\,.
  \eql{trans4}
  $$
The first equality means that the previous scalar analysis holds unchanged for $F$ and the
revival attenuation is again $1/m^{d+1}$, for $m$ even. The same argument applies to
$Y(q)$. Shifting the origin to $s=1$, $Y(q)$ has revivals at $1/m'$ where $m'$ is even.
$m$ and $m'$ are related by $m'=m-1$, because of the shift. Hence from (\peq{trans4}),
$F$ will exhibit sign reversed revivals of $Y$ at $s=1/m$ for $m$ odd. The attenuation for
these two series of revivals is $1/m^{d+1}$ from $F(s=0) $ and $F(s=1)$

The relevant ratio in sizes between these two sets of revivals is given by the ratio of the
corresponding Stefan--Boltzmann constants at the two starting points, $s=0,1$, \ie,
  $$
(1-2^{-2l+1})\,.
  $$

For $l=2$, which is a good approximation for the three--sphere, this ratio equals $7/8$
which can be verified numerically from the complete expression for $\Xi'\sim \log Z$, also
plotted in Fig.3.

These results are for the partial {\it energies}, but, for any $d$, the complete expressions
are available for the partition function, $\Xi'$, (\peq{ssum}) and (\peq{ssf}), for normal
bosons and fermions. Then, for example, the normal spinor can be translated from $s=0$
to $s=1$ to give,

$$\eqalign{
  -\sum_{m=1}^\infty(-1)^m{1\over m}\,\cosech^d(m(\be+2\pi i)/2)
  =-\sum_{m=1}^\infty(-1)^m{1\over m}(-1)^{md}\cosech^d(m\be/2)\cr
  =-\sum_{m=1}^\infty{1\over m}\cosech^d(m\be/2)\,,\quad d \,\,{\rm odd}
  }
  \eql{ssft}
  $$
and this is minus the $\Xi'$ for a spinor boson in agreement with (\peq{trans}), for the
energy.

If $d$ is even, the utility of partial quantities disappears. Only the complete expressions
are available, and then just for translations. For scalars, one has,

$$\eqalign{
  \Xi_d'(\be+2\pi i)={1\over 2^d}\sum_{m=1}^\infty{(-1)^{m(d+1)}
  \over m}\cosh (m\be/2)
  \,\cosech^d (m\be/2)\,,
  }
  \eql{ssumt}
  $$

Hence if $d$ is even,
$$\eqalign{
  \Xi_d'(\be+2\pi i)={1\over 2^d}\sum_{m=1}^\infty{(-1)^m
  \over m}\cosh (m\be/2)
  \,\cosech^d (m\be/2)\,,
  }
  \eql{ssumte}
  $$
which is the negative of a fermionic scalar.

In this way, the integer sign reversed revivals, in, say, Fig.1 can be understood, as noted
in [\pref{Cardy2}] (obtained slightly differently).

Glaisher also gives the power spectra of these two `systems'. That for the bosonic spinor
is the same as that for the normal spinor, Fig.6. The fermionic scalar spectrum  for $l=2$
is given in Fig.8.

\epsfxsize=5truein \epsfbox{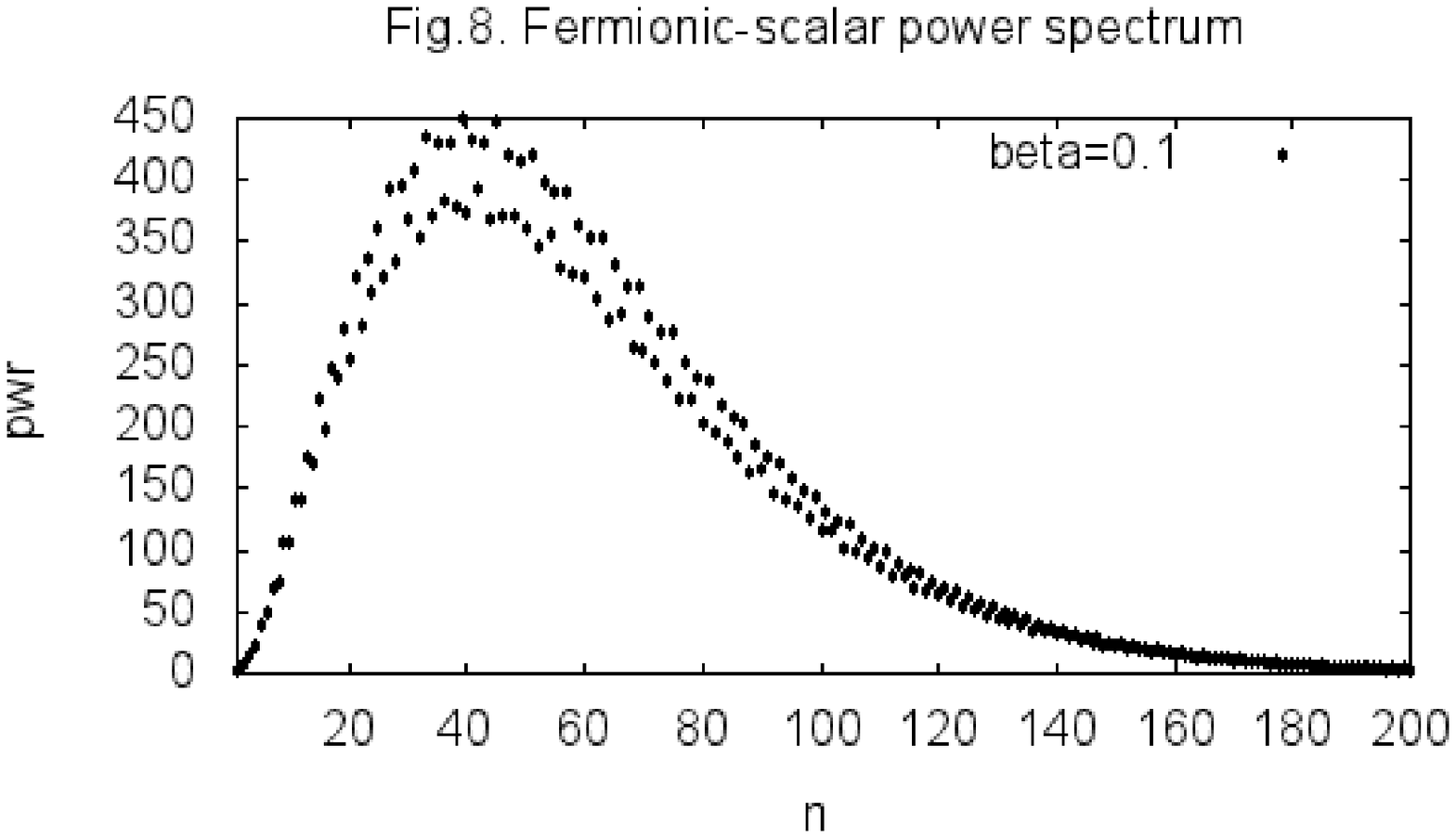}

I recast some of the above more abstractly in terms of the $S$ and $T$ generators of the
modular group. For this purpose, it is algebraically convenient to absorb the factors of
$\xi^2$ in (\peq{inv}) by defining
  $$
  \overline \ep_l(\xi)={1\over\xi^2}\,\ep_l(\xi)
  $$
so that $\ol\ep_l(1/\xi)=(-1)^l\,\ol\ep_l(\xi)$, and likewise for $X_l, Y_l, F_l$ and $B_l$.

The actions of the modular group can be summarised as,
    $$
          {\bf S}\left(\matrix{\ol F\cr \ol Y\cr \ol X}\right)_{\!\!l}=
          \left(\matrix{S\ol F\cr S\ol Y\cr S\ol X}\right)_{\!\!l}=
          (-1)^l\,\left(\matrix{\ol F\cr -\ol X\cr -\ol Y}\right)_{\!\!l}
    $$
and
  $$
          {\bf T}\left(\matrix{\ol F\cr \ol Y\cr \ol X}\right)_{\!\!l}=
          \left(\matrix{T\ol F\cr T\ol Y\cr T\ol X}\right)_{\!\!l}=
          (-1)^l\,\left(\matrix{-\ol Y\cr -\ol F\cr \ol X}\right)_{\!\!l}\,,
    $$
so that the matrix  representations are,
     $$
      {\bf S}=(-1)^l\left(\matrix{1&0&0\cr0&0&-1\cr0&-1&0}\right)\,,\quad
      {\bf T}=(-1)^l\left(\matrix{0&-1&0\cr-1&0&0\cr0&0&1}\right)\,.
     $$
I separate $\ol B$ as it does not mix.

It is seen that, in addition to the usual relations, (\viz\ ${\bf S}^2={\bf 1}$,
$(\bf{ST})^3={\bf 1}$) one has ${\bf T}^2={\bf 1}$. I note that $X=ST\,F$.

The behaviour around a rational $s=p_1/p_2$ can be obtained from the modular
transformation connecting the origin to the point $s=p_1/p_2$. This is achieved, as said in
[\pref{Cardy1}], using the transformation $T^{a_1}S\, T^{a_2}S\ldots$ where the $a_i$
are the integers in the continued fraction form of $p_1/p_2$. (The order of the operators is
such as to take $s=p_1/p_2$ back to $s=0$.) It is clear from the above actions that one
gets either $F$, $-Y$ or $X$, depending on the particular rational. The behaviour around
any rational, if it can be discerned, could therefore be considered as an attenuated revival
of one of these three functions, evaluated at its origin. Visible examples in Fig.5 are at
$s=1/2$ giving $F$, at $s=1/3$ giving $-Y$ and at $s=2/3$ giving $X$.\footnote{ I am
taking the same connection between the energy and partition function that allows
(\peq{mod2}) to be derived.}

I finally note that $B,F,X$ and $Y$ correspond to the set of elliptic functions, $zs,ds,cs$
and $ns$, which are grouped together by Glaisher, [\pref{Glaisher2}].
\section{\bf 8. Discussion and conclusion}

This paper is, essentially, just a commentary on a recent work of Cardy, [\pref{Cardy2}],
concerning quantum revivals in higher dimensions, with some additions. The revivals at
the rationals can be explained, for odd dimensional spheres, by modular behaviour. There
is no such exact property for even spheres but rational  revivals can still be detected.

As a small novelty, the calculations have been extended to the fermion field.

It is also shown, incidentally, that the entanglement entropy does not equal the
thermodynamic one, except for $d=1$, the torus.

\section{\bf Appendix A. The partition function, an observation}

It is well known that there are two equivalent ways of evaluating the partition function.
One relies on the CFT operator counting method (in flat space) and the other on solving
the energy eigenvalue problem on the spatial section of the conformally related, curved
manifold. Here, I would like to make a few technical remarks on this equivalence in the
present set up, restricting myself to the scalar, boson field.

The most appropriate form of the operator counting method for me is that outlined in
[\pref{Cardy3}] and [\pref{Cardy2}]. I have to repeat some of this known material in
order to make my point. See also Kutasov and Larsen, [\pref{KandL}].

Because of the conformal relation, the time translation generator on the (Euclidean)
Einstein universe, S$^1\times$S$^d$, is proportional to the scale generator on the flat
$\oR^{d+1}$. This implies that the energy of a state, $E$, equals the scaling dimension
of the corresponding field which means that the partition function (actually just the sum
over states part) is the generating function for the complete set of modular weights,
$\De$, \ie,
  $$
  \Xi'(\be)=\sum_\De e^{-\be\De}\,.
  $$

The total set of independent operators in $\oR^{d+1}$ is,
  $$
  \prod_j \,\prod_{i=1}^{d+1}\pa_i^{n_i^{(j)}}\phi\,,
  \eql{op}
  $$
modulo the equation of motion, $\pa_i\pa^i\,\phi=0$, a requirement that can be
implemented by restricting $n^{(j)}_{d+1}$ to $0$ and $1$, as explained in
[\pref{Cardy3,Cardy2}]. This leaves $n_1\ldots n_d$ as {\it unrestricted}, non-negative
integers.

The scaling dimension of the operator, (\peq{op}), then splits into two,
  $$
  \De=\sum_j\big((d-1)/2+ n_1^{(j)}+\ldots+n_d^{(j)}\big)+
   \sum_j\big((d+1)/2+ n_1^{(j)}+\ldots+n_d^{(j)}\big)\,,
   \eql{delta}
  $$
and the partition function factorises, implying,
  $$
  \Xi'_d(\be)=\sum_{{\bf n=0}}^\infty{1\over1-\exp{-\be(a_N+{\bf n.1})}}+
  \sum_{{\bf n=0}}^\infty{1\over1-\exp{-\be(a_D+{\bf n.1})}}\,,
  \eql{logs}
  $$
with $a_N=(d-1)/2,\,\,\,a_D=a_N+1$.

The significance of this is that the quantities $a_{N,D}+{\bf n.1}$ are recognised as the
conformal single particle energies (the eigenvalues of $\sqrt D$) on a $d$--hemisphere
with Neumann and Dirichlet conditions on the rim.\footnote{ It is interesting to remark
that the pseudo--operator, $H_d=\sqrt{D_d}$, on the $d$--sphere can be defined
recursively by $H_d=L\oplus H_{d-1}$ with $L$ realised as the operator $-id/d\phi$ acting
on $\oZ_2$ twisted fields on the circle. $L$ has eigenvalues $n+1/2$ and iterating from
$H_0=\mp1/2$ gives the eigenvalues on the $N,D$ hemisphere mentioned above.
$H_0^2=1/4$ means that the Yamabe--Penrose operator vanishes in dimension 0,
consonant with its form.} Uniting these gives the full sphere result and so the equivalence
of the two approaches has been verified with no work. There is no need to perform the
combinatorics in say (\peq{delta}) in order to obtain the degeneracies of the eigenlevels,
nor any group theory likewise (in this simple case).

Equation (\peq{eff}) provides an alternative to (\peq{logs}) and, after very slight algebra,
reads,
  $$
  \Xi'_d(\be)={1\over2^d}\sum_{m=1}^\infty{1\over m}\,\cosh(m\be/2)\,
  \cosech^d(m\be/2)\,,
  \eql{sssd}
  $$
as used above,  (\peq{ssum}). Expression (\peq{sssd}) is derived in [\pref{ChandD}]
\footnote{ The spatial geometry was, more generally, an orbifolded sphere.} and also in
[\pref{Apps,DandA1,DandA2}].

This particular equivalence of approaches has been discussed recently in some detail by
Beccaria, Bekaert and Tseytlin, [\pref{BBT}], who employ degeneracies and apply the
harmonic condition, $\pa_i\pa^i\,\phi=0$, differently, so that there is no split,
(\peq{logs}). They retrieve (\peq{sssd}).

\section{\bf Appendix B. Entanglement entropy}

While it is in view, the dependence on $B$ given in the section 6 allows an entanglement
entropy to be evaluated. $B$ is the inverse of the replica covering, and there is a conical
singularity on the manifold of extent S$^1\times$S$^{d-1}$ which codimension 1
manifold forms the entangling surface. I give a few details.

According to the usual rule, the derivative of $\Xi$ with respect to $B$ at $B=1$ is
required. This formulae is
  $$
  S_E=-\big(1+B\pa _B\big)\, \Xi(B)\bigg|_{B=1}\,.
  \eql{rule}
  $$

First, for bosons, the sum part gives,
  $$\eqalign{
  {\pa\over \pa B}\Xi'_d(\be,B)\bigg|_{B=1}=-{1\over2^{d}}
  \sum_{m=1}^\infty{1\over m}{m\be\over2}\cosech^{d+1}(m\be/2)\cr
  }
  $$
and
  $$
  \Xi'_d(\be,1)={1\over2^{d}}
  \sum_{m=1}^\infty{1\over m}\coth(m\be/2)\,\cosech^{d-1}(m\be/2)\,,
  $$
so the sum contribution to the entanglement entropy $S$ is
  $$\eqalign{
  &{1\over2^{d}}
  \sum_{m=1}^\infty{1\over m}\bigg({m\be\over2}\cosech^2(m\be/2)+\coth(m\be/2)
  \bigg)\cosech^{d-1}(m\be/2)\cr
  &={1\over2^{d}}
  \sum_{m=1}^\infty{1\over m}\bigg({m\be\over2}\cosech(m\be/2)+\cosh(m\be/2)\,.
  \bigg)\cosech^{d}(m\be/2)\,.\cr
  }
  $$

The vacuum energy on a lune of angle $2\pi/B$ is given in [\pref{ChandD}] (see also
[\pref{Dowpist}] )as a generalised Bernoulli polynomial,

$$\eqalign{
  E_0(1)=-B^{(d)}_{d+1}\big((d+1)/2\mid{\bf1}\big)\cr
 ( \pa/\pa B)E_0(B)\big|_{B=1}=-B^{(d+1)}_{d+1}\big((d+1)/2\mid{\bf1}\big)\,.\cr
  }
  $$

Turning now to the fermion case, the eigenvalues of the propagating Dirac operator on the
$d$--dimensional periodic $B$--lune are again perfect squares for conformal in $d+1$
dimensions,
  $$
  \la_{\bf n}(B)=\big(a+{\bf n.}\bom\big)^2\,\quad {\bf n}={\bf 0}\ldots{\bf\infty}\,,
  \eql{speigs}
  $$
where the $d$--vector $\bom$ equals $(B,{\bf 1})$ and $a=(d-1+B)/2$. The corresponding
cylinder kernel is then
  $$\eqalign{
  K_B^{1/2}(\tau)&=\sum_{\bf n=0}^\infty e^{-(a+{\bf n.}\bom)\tau}\cr
  &={1\over2^d}\,\prod_{i=1}^d\,\cosech(\om_i\tau/2)\cr
  &={1\over2^d}\,\cosech(B\tau/2)\,\cosech^{d-1}(\tau/2)\,.
  }
  \eql{ckf}
  $$

By general statistical mechanics, the fermion  free energy reads,
  $$
  -\be F=\Xi^f=-\be E^f_0(B)+\sum_{m=1}^\infty(-1)^m {1\over m}\,K^{1/2}_B(m\be)\,,
  \eql{ffe}
  $$
where $E^f_0(B)$ is the fermion Casimir energy on the lune.

The entanglement entropy entails a derivative with respect to $B$ and the sum
contribution involves
  $$
  {\pa\over\pa B}\, K_B^{1/2}(m\be)\bigg|_{B=1}
  =-{m\be\over2^{d+1}}\coth(m\be/2)\,\cosech^{d}(m\be/2)
  $$
and
$$
   K^{1/2}_1(m\be)={1\over2^d}\cosech^d(m\be/2)\,.
  $$

Hence from (\peq{rule}) the corresponding contribution to the entanglement entropy is
  $$
  S^f_2={1\over2^d}\sum_{m=1}^\infty{(-1)^m\over m}\,\cosech^d(m\be/2)
  \bigg(1-{m\be\over2}\coth(m\be/2)\bigg)\,.
  $$

To complete the evaluation, the Casimir term part is required. This is linear in $\be$,

  $$\eqalign{
   S^f_1&=-\beta\big(E_0(1)+\pa/\pa B\,E_0(B)\big|_{B=1}\big)\cr
   &\equiv A^f_d\,\be
   }
  $$
and can be deduced from the relevant  \zf\ which is again a Barnes one and, drawing upon
[\pref{Dowpist}], for odd $d$, this time,
  $$\eqalign{
  E^f_0(1)=B^{(d)}_{d+1}\big(d/2+1\mid{\bf1}\big)\cr
 ( \pa/\pa B)E^f_0(B)\big|_{B=1}=B^{(d+1)}_{d+1}\big(d/2+1\mid{\bf1}\big)\,.\cr
  }
  $$
These numbers are easily computed $d$ by $d$.

For $d=1$, similar expressions (actually modularly transformed) are derived in
[\pref{ANT}] in a  somewhat more involved way using twist operators.

It is interesting to compare the entanglement and the thermodynamical entropy, which is
defined by $S_T=\big(1-\be(\pa/\pa\be) \big)\Xi(\be,1)$. It is easy to show that they are
not equal, unless $d=1$.

\begin{ignore}

CHECKING:

Look at (\peq{ckf}) and consider the log form of ({\peq{ffe}). FD

  $$\eqalign{
  \be F= -\sum_{\bf n}\log(1+e^{-(a+{\bf n.1})\be}\big)&=
  \sum_{\bf n}\sum_{m=1}^\infty{(-1)^m\over m}
  e^{-(a+{\bf n.1})m\be}=\!\sum_{m=1}^\infty\!{(-1)^m\over m} K^{(1/2}(m\be)\cr
  &={1\over2^d}\!\sum_{m=1}^\infty\!{(-1)^m\over m}\cosech^d(m\be/2)
  }
  $$

$\be F=-\log Z$

Check basic definitions.

BE:

$$
  F=-L
$$


LandL' have for the thermodynamic potential,

  $$
  \be F=\sum\log(1-e^{-\om\be})
  $$
which is BE free energy, or thermodynamic potential, times $\be$, (Landau and Lifshits
(53.4))

Apps p16, 18

  $$
 -\log Z\equiv W={1\over2}\log\det \De=-{1\over2}\ze'(0)
  $$
therefore $W=\be F$

Cardy has
$$
\log Z=-{1\over2}\log\det(-\nabla^2+)
$$

Cardy plots $\log Z$ \ie\ $-W$ or $-\be F$.

In Klebanov one has $F_K\equiv -\log Z=\log\det(-\nabla^2+\ldots)$

Apps p105 computes scalar W

\eg\ involves
  $$
-W\sim\sum {1\over m} \cosh\times \ldots
  $$
or
  $$
-\be F\sim\sum {1\over m} \cosh\times \ldots
  $$
as finite temperature correction to $-\be F$ which is what is plotted.

Now have to check Chang and Dowker (78) last term. This is a standard formula.
  $$
  K^{(1/2)}(\tau)\equiv\sum_{\om}e^{-\om\tau}
  $$
$\om $ is single particle energy.

From LandL', again, BE,
$$
  \be F=\sum_{\om}\log(1-e^{-\om\be})=-\sum_{\om}\sum_{m=1}^\infty {1\over m}
  e^{-\be m}=-\sum_m {1\over m} K^{(1/2)}(\be m)
  $$
which agrees with ChandD.

For FERMIONS, LandL' (52.4),

  $$
  -\be F=\sum_{\om}\log(1+e^{-\om\be})=-\sum_{\om}\sum_{m=1}^\infty {(-1)^m\over m}
  e^{-\be m\om}=-\sum_m {(-1)^{m}\over m} K^{(1/2)}(\be m)
  $$

Let's evaluate on sphere/cylinder as in QRCFT (24)
  $$
  K^{(1/2})(\tau)\sim \cosech
  $$
implies
  $$
  -\be F=-\sum_m {(-1)^{m}\over m} \cosech
  $$
which should be plotted.

 Apps p65
  $$
  W={1\over2}\ze'(0)
  $$

He has
  $$
  W\sim\sum{(-1)^m\over m} \cosech
  $$

 ******************************************************

 CHECK ALGEBRA

Start from (\peq{sinv}). Use $E'=(\pa/\pa \be)(\be F')=-(\pa/\pa \be)\,\log Z'$ for finite
temperature corrections. Set $a=1$ then $E'=\ep'(\xi)$ and assume $l$ understood.
$\be=2\pi/\xi$. So
  $$
  \ep'(\xi)=(2\pi)^{-1}\,\xi^2{\pa\over \pa\xi}\,\lg'(\xi)
  $$
and therefore
  $$\eqalign{
  \ep'(1/\xi)&=(2\pi)^{-1},\xi^{-2}{\pa\over \pa1/\xi}\,\lg'(1/\xi)\cr
  &=(2\pi)^{-1}\,\xi^{-2}(-\xi^2){\pa\over \pa\xi}\,\lg'(1/\xi)\cr
  &=-(2\pi)^{-1}\,{\pa\over \pa\xi}\,\lg'(1/\xi)\cr
  }
  $$
 Then (\peq{sinv}) becomes,
  $$
  e(0)+\ep'(\xi)=
  (-1)^l\xi^{2l}\big(\ep(0)+\ep'(1/\xi)\big)\,.
  \eql{sinv}
  $$

Therefore

$$
\ep'(\xi)=(2\pi)^{-1}\,\xi^2{\pa\over \pa\xi}\,\lg'(\xi)=-(-1)^l\xi^{2l}
(2\pi)^{-1}\,{\pa\over \pa\xi}\,\lg'(1/\xi)+\ep(0)\big((-1)^l\xi^{2l}-1\big)
$$

Therefore
  $$\eqalign{
{\pa\over \pa\xi}\,\lg'(\xi)&=-(-1)^l\xi^{2l-2}
\,{\pa\over \pa\xi}\,\lg'(1/\xi)+2\pi\ep(0)\big((-1)^l\xi^{2l}-1\big)\xi^{-2}\cr
&=
}
$$
\end{ignore}

\newpage
 \vglue 20truept
 \noin{\bf References.} \vskip5truept
\begin{putreferences}
   \ref{DandKi}{Dowker,J.S. and Kirsten,K. {\it Elliptic functions and temperature inversion on
   spheres}. \np{638}{2002}{405}.}
   \ref{Dowfint}{Dowker,J.S. {\it Finite temperature and vacuum effects in higher dimensions},
   \break \cqg{1}{1984}{359}.}
   \ref{Berndt}{Berndt,B.C. {\it Analytic Eisenstein Series, Theta Functions and series
   relations in the spirit of Ramanujan}, \jram{303/304}{1978}{332}.}
   \ref{CapandC}{Cappelli,A. and Costa,A. {\it On the stress tensor of conformal field theories
   in higher dimensions}, \np{314}{1989}{707}.}
   \ref{GPP}{Gibbons,G.W., Perry,M.J. and Pope,C.N. {\it Partition Functions, the
   Bekenstein Bound and Temperature Inversion in Anti--de Sitter Space and its Conformal
   Boundary}, \prD {74} {2006} 084009. }
   \ref{Dowzerom}{Dowker,J.S., {\it Zero modes, entropy bounds and
   partition functions}, \break \cqg {20}{2003} {L105}.}
   \ref{Dowpist}{Dowker,J.S. {\it Spherical Casimir pistons}, \cqg{28}{2011}{155018}.}
   \ref{KandL}{Kutasov D. and  Larsen,F. {\it Partition Sums and Entropy Bounds in Weakly
   Coupled CFT}, {\it JHEP}  0101:001,2001.}
   \ref{DandKi2}{Dowker,J.S. and Kirsten,K. {\it Elliptic aspects of statistical mechanics on
   spheres}, \jmp {49}{2008}{113513}.}
   \ref{ANT}{Azeyanagi,T., Nishioka,T. and Takayanagi,T. {\it Near extremal black hole entropy
   as entanglement entropy via $AdS_2/CFT_1$}, \prD {7}  {2008}{064005}.}
   \ref{BBT}{Beccaria,M., Bekaert,X. and Tseytlin,A.A. {\it Partition function of free
   conformal higher spin theory}, {\it JHEP} {\bf 08}(2014)113 }
   \ref{AandD}{Altaie,M.B. and Dowker,J.S. {\it Spinor fields in an Einstein universe:
   Finite temperature effects},\prD{18}{1978}{3557}.}
   \ref{Glaisher2}{Glaisher,J.W.L. {\it On the series which represent the twelve elliptic
   and four zeta functions}, {\it Messenger Math.} {\bf 18} (1889) 1.}
   \ref{Glaisher1}{Glaisher,J.W.L. {\it On certain sums of products of quantities
   depending on the divisors of a number}, {\it Messenger Math.} {\bf 15} (1886) 1.}
   \ref{Dowmod}{Dowker,J.S. {\it Modular properties of Eisenstein series and statistical
   physics.} \break ArXiv:0810.0537}
   \ref{DandA1}{Dowker,J.S. and Apps,J.S., {\it Further functional determinants},
   \cqg{12}{1995}{1363}; ArXiv:hep-th/9502015.}
    \ref{DandA2}{Dowker,J.S. and Apps,J.S., {\it Functional determinants on certain
    domains}, {\it Int. J.Mod.Phys.}{\bf 5} (1996) 799. ArXiv:hep-th/9506205}
   \ref{Unwin1}{Unwin,S.D. {\it Selected quantum field theory effects in multiply
 connected spacetimes}. Thesis, University of Manchester, 1980.}
 \ref{Unwin2}{Unwin,S.D. \jpa{13}{1980}{313}.}
  \ref{DandK}{Dowker,J.S. and Kennedy,G. {\it Finite temperature and boundary effects in
  static space--times}, \jpa{11}{1978}{895}.}
 \ref{Kennedy}{Kennedy,G. {\it Topological symmetry restoration}, \prD{23}{1981}{2884}.}
    \ref{Cardy2}{Cardy,J. {\it Quantum revivals in Conformal Field Theories in Higher
    Dimensions}, ArXiv:1603.08267}
     \ref{Cardy1}{Cardy,J. {\it Thermalization and Revivals after a Quantum
     Quench in Conformal Field Theory}, \prl{112}{2014}{220401}.}
     \ref{Cardy3}{Cardy,J. {\it Operator content and modular properties of higher dimensional
     conformal field theories}, \np{366}{1991}{403}.}
    \ref{Barnesa}{Barnes,E.W. {\it Trans. Camb. Phil. Soc.} {\bf 19} (1903)
  374.}
  \ref{Sch1}{Schr\"odinger, E.W. {\it Expanding Universes} (C.U.P. 1956. Cambridge).}
   \ref{Sch2}{Schr\"odinger, E.W. {\it Proc. Roy. Irish Acad.} {\bf 46A} (1946) 25.}
   \ref{Wenger}{Wenger,D.L. \jmp{8}{1967}{135}.}
  \ref{Barnesb}{E.W.Barnes {\it Trans. Camb. Phil. Soc.} {\bf 19} (1903)
  426.}
  \ref{Elizalde}{Elizalde,E. {\it Math. of Comp.} {\bf 47} (1986) 347.}
  \ref{doweven}{Dowker,J.S. {\it Entanglement entropy on even spheres} ArXiv:1009.3854.}
  \ref{CandT}{Copeland,E. and Toms,D.J. \np {255}{1985}{201}.}
  \ref{Dowmasssphere}{Dowker,J.S. {\it Massive sphere determinants} ArXiv:1404.0986.}
  \ref{dowtwist}{Dowker,J.S. {\it Conformal weights of charged R\'enyi entropy
  twist operators for free scalars in arbitrary dimensions.} ArXiv:1508.02949.}
   \ref{Dowlensmatvec}{Dowker,J.S. {\it Lens space matter determinants in the vector
   model},  ArXiv:\break 1405.7646.}
   \ref{dowtwist}{Dowker,J.S. {\it Conformal weights of charged R\'enyi entropy twist
   operators for free scalar fields in arbitrary dimensions} ArXiv:1509.00782.}
   \ref{GandS}{Gel'fand,I.M. and Shilov,G.E. {\it Generalised Functions} Vol.1 (Academic Press,
   New York, 1964.}
   \ref{BandS}{Bogoliubov,N.N. and Shirkov,D.V. {\it Introduction to the theory of quantized
   fields} (Interscience, New York, 1959.).}
   \ref{dowsignch}{Dowker,J.S. \jpa{2}{1969}{267}.}
   \ref{MandD}{Dowker,J.S. and Mansour,T. {\it J.Geom. and Physics} {\bf 97} (2015) 51.}
   \ref{dowaustin}{Dowker,J.S. 1979 {\it Selected topics in topology and quantum
    field theory}
    \ref{Dowmultc}{Dowker,J.S. \jpa{5}{1972}{936}.}
    (Lectures at Center for Relativity, University of Texas, Austin).}
   \ref{Dowrenexp}{Dowker,J.S. {\it Expansion of R\'enyi entropy for free scalar fields}
   ArXiv:1408.0549.}
   \ref{EandH}{Elvang, H and  Hadjiantonis,M.  {\it Exact results for corner contributions to
   the entanglement entropy and R\'enyi entropies of free bosons and fermions in 3d} ArXiv:
   1506.06729 .}
   \ref{SandS}{T.Souradeep and V.Sahni \prD {46} {1992} {1616}.}
   \ref{CandH}{Casini H., and Huerta,M. \jpa{42}{2009}{504007}.}
   \ref{CandH2}{Casini H., and Huerta,M. J.Stat.Mech {\bf 0512} (2005) 12012.}
   \ref{CandC}{Cardy,J. and Calabrese,P. \jpa{42}{2009}{504005}.}
   \ref{CaandH}{Casini,H. and Huerta,M. \plb{694}{2010}{167}.}
    \ref{Dow7}{Dowker,J.S. \jpa{25}{1992}{2641}.}
    \ref{Dowcosecs}{Dowker,J.S. {\it On sums of powers of cosecs}, ArXiv:1507.01848.}
    \ref{Jeffery}{Jeffery, H.M. \qjm{6}{1864}{82}.}
   \ref{BMW}{Bueno,P., Myers,R.C. and Witczak--Krempa,W. {\it Universal corner entanglement
   from twist operators} ArXiv:1507.06997.}
  \ref{BMW2}{Bueno,P., Myers,R.C. and Witczak--Krempa,W. {\it Universality of corner
  entanglement in conformal field theories} ArXiv:1505.04804.}
  \ref{BandM}{Bueno,P., Myers,R.C. {\it Universal entanglement for higher dimensional
  cones} ArXiv:1508.00587.}
   \ref{Dowstat}{Dowker,J.S. \jpa{18}{1985}3521.}
   \ref{Dowstring}{Dowker,J.S. {\it Quantum field theory around conical defects} in {\it
   The Formation and Evolution of Cosmic Strings} edited by Gibbons,G.W, Hawking,S.W. and
   Vachaspati,T. (CUP, Cambridge, 1990).}
   \ref{Hung}{Hung,L-Y.,Myers,R.C. and Smolkin,M. {\it JHEP} {\bf 10} (2014) 178.}
\ref{Dow7}{Dowker,J.S. \jpa{25}{1992}{2641}.}
   \ref{B}{Belin,A.,Hung,L-Y., Maloney,A., Matsuura,S., Myers,R.C. and Sierens,T.\break
   {\it JHEP} {\bf 12} (2013) 059.}
   \ref{B2}{Belin,A.,Hung,L-Y.,Maloney,A. and Matsuura,S.
   {\it JHEP01} (2015) 059.}
   \ref{Norlund}{N\"orlund,N.E. \am{43}{1922}{121}.}
    \ref{Norlund1}{N\"orlund,N.E. {\it Differenzenrechnung} (Springer--Verlag, 1924, Berlin.)}
   \ref{Dowconearb}{Dowker,J.S. \prD{36}{1987}{3742}.}
     \ref{Dowren}{Dowker,J.S. \jpamt {46}{2013}{2254}.}
     \ref{DandB}{Dowker,J.S. and Banach,R. \jpa{11}{1978}{2255}.}
     \ref{Dowcen}{Dowker,J.S. {\it Central Differences, Euler numbers and
   symbolic methods} ArXiv: 1305.0500.}
   \ref{Dowcone}{Dowker,J.S. \jpa{10}{1977}{115}.}
   \ref{schulman2}{Schulman,L.S. \jmp{12}{1971}{304}.}
   \ref{DandC}{Dowker,J.S. and Critchley,R. {\it Vacuum stress tensor in an Einstein
   universe: Finite temperature effects},  \prD{15}{1977}{1484}.}
     \ref{Thiele}{Thiele,T.N. {\it Interpolationsrechnung} (Teubner, Leipzig, 1909).}
     \ref{Steffensen}{Steffensen,J.F. {\it Interpolation}, (Williams and Wilkins,
    Baltimore, 1927).}
     \ref{Riordan}{Riordan,J. {\it Combinatorial Identities} (Wiley, New York, 1968).}
     \ref{BSSV}{Butzer,P.L., Schmidt,M., Stark,E.L. and Vogt,I. {\it Numer.Funct.Anal.Optim.}
    {\bf 10} (1989) 419.}
      \ref{Dowcascone}{Dowker,J.S. \prD{36}{1987}{3095}.}
      \ref{Stern}{Stern,W. \jram {79}{1875}{67}.}
     \ref{Milgram}{Milgram, M.S., Journ. Maths. (Hindawi) 2013 (2013) 181724.}
     \ref{Perlmutter}{Perlmutter,E. {\it A universal feature of CFT R\'enyi entropy}
     ArXiv:1308.1083 }
     \ref{HMS}{Hung,L.Y., Myers,R.C. and Smolkin,M. {\it Twist operators in
     higher dimensions} ArXiv:1407.6429.}
     \ref{ABD}{Aros,R., Bugini,F. and Diaz,D.E. {\it On the Renyi entropy for
     free conformal fields: holographic and $q$--analog recipes}.ArXiv:1408.1931.}
     \ref{LLPS}{Lee,J., Lewkowicz,A., Perlmutter,E. and Safdi,B.R.{\it R\'enyi entropy.
     stationarity and entanglement of the conformal scalar} ArXiv:1407.7816.}
     \ref{Apps}{Apps,J.S. {\it The effective action on a curved space and its conformal
     properties} PhD thesis (University of Manchester, 1996).}
   \ref{CandD}{Candelas,P. and Dowker,J.S. {\it Field theories on conformally
   related space-times: Some global considerations}, \prD{19}{1979}{2902}.}
    \ref{Hertzberg}{Hertzberg,M.P. \jpa{46}{2013}{015402}.}
     \ref{CaandW}{Callan,C.G. and Wilczek,F. \plb{333}{1994}{55}.}
    \ref{CaandH}{Casini,H. and Huerta,M. \plb{694}{2010}{167}.}
    \ref{Lindelof}{Lindel\"of,E. {\it Le Calcul des Residues} (Gauthier--Villars, Paris,1904).}
    \ref{CaandC}{Calabrese,P. and Cardy,J. {\it J.Stat.Phys.} {\bf 0406} (2004) 002.}
    \ref{MFS}{Metlitski,M.A., Fuertes,C.A. and Sachdev,S. \prB{80}{2009}{115122}.}
    \ref{Gromes}{Gromes, D. \mz{94}{1966}{110}.}
    \ref{Pockels}{Pockels, F. {\it \"Uber die Differentialgleichung $\De
  u+k^2u=0$} (Teubner, Leipzig. 1891).}
   \ref{Diaz}{Diaz,D.E. JHEP {\bf 7} (2008)103.}
  \ref{Minak}{Minakshisundaram,S. {\it J. Ind. Math. Soc.} {\bf 13} (1949) 41.}
    \ref{CaandWe}{Candelas,P. and Weinberg,S. \np{237}{1984}{397}.}
     \ref{Chodos1}{Chodos,A. and Myers,E. \aop{156}{1984}{412}.}
     \ref{ChandD}{Chang,P. and Dowker,J.S. {\it Vacuum energy on orbifold factors of spheres}, \np{395}{1993}{407}.}
    \ref{LMS}{Lewkowycz,A., Myers,R.C. and Smolkin,M. {\it Observations on
    entanglement entropy in massive QFTs.} ArXiv:1210.6858.}
    \ref{Bierens}{Bierens de Haan,D. {\it Nouvelles tables d'int\'egrales
  d\'efinies}, (P.Engels, Leiden, 1867).}
    \ref{DowGJMS}{Dowker,J.S.  \jpa{44}{2011}{115402}.}
    \ref{Doweven}{Dowker,J.S. {\it Entanglement entropy on even spheres.}
    ArXiv:1009.3854.}
     \ref{Dowodd}{Dowker,J.S. {\it Entanglement entropy on odd spheres.}
     ArXiv:1012.1548.}
    \ref{DeWitt}{DeWitt,B.S. {\it Quantum gravity: the new synthesis} in
    {\it General Relativity} edited by S.W.Hawking and W.Israel (CUP,Cambridge,1979).}
    \ref{Nielsen}{Nielsen,N. {\it Handbuch der Theorie von Gammafunktion}
    (Teubner,Leipzig,1906).}
    \ref{KPSS}{Klebanov,I.R., Pufu,S.S., Sachdev,S. and Safdi,B.R.
    {\it JHEP} 1204 (2012) 074.}
    \ref{KPS2}{Klebanov,I.R., Pufu,S.S. and Safdi,B.R. {\it F-Theorem without
    Supersymmetry} 1105.4598.}
    \ref{KNPS}{Klebanov,I.R., Nishioka,T, Pufu,S.S. and Safdi,B.R. {\it Is Renormalized
     Entanglement Entropy Stationary at RG Fixed Points?} 1207.3360.}
    \ref{Stern}{Stern,W. \jram {79}{1875}{67}.}
    \ref{Gregory}{Gregory, D.F. {\it Examples of the processes of the Differential
    and Integral Calculus} 2nd. Edn (Deighton,Cambridge,1847).}
    \ref{MyandS}{Myers,R.C. and Sinha, A. \prD{82}{2010}{046006}.}
   \ref{RyandT}{Ryu,S. and Takayanagi,T. JHEP {\bf 0608}(2006)045.}
    \ref{Dowcmp}{Dowker,J.S. \cmp{162}{1994}{633}.}
     \ref{Dowjmp}{Dowker,J.S. \jmp{35}{1994}{4989}.}
      \ref{Dowhyp}{Dowker,J.S. \jpa{43}{2010}{445402}.}
       \ref{HandW}{Hertzberg,M.P. and Wilczek,F. \prl{106}{2011}{050404}.}
      \ref{dowkerfp}{Dowker,J.S.\prD{50}{1994}{6369}.}
       \ref{Fursaev}{Fursaev,D.V. \plb{334}{1994}{53}.}
\end{putreferences}

\bye